\documentclass[12pt]{article}

\usepackage{scicite}

\usepackage{times}
\usepackage{amsmath}%
\usepackage{graphicx}%

\usepackage{hyperref}
\hypersetup{colorlinks=true,linkcolor=blue,linktocpage,citecolor=blue,urlcolor=blue}

\makeatletter
\g@addto@macro{\UrlBreaks}{\UrlOrds}
\makeatother

\newenvironment{sciabstract}{%
\begin{quote} \bf}
{\end{quote}}

\title{Anthropogenic contributions to slow warming over 1998-2012}

\author{Xuanming Su$^{1,\ast}$, Hideo Shiogama$^{2}$, Katsumasa Tanaka$^{2,3}$, \\Kaoru Tachiiri$^{4}$, Tomohiro Hajima$^{4}$, Michio Watanabe$^{4}$, \\Michio Kawamiya$^{4}$, Kiyoshi Takahashi$^{1}$ \& Tokuta Yokohata$^{2}$ \\ 
\normalsize{$^{1}$Social Systems Division, National Institute for Environmental} \\ 
\normalsize{Studies(NIES), Tsukuba, Japan}\\
\normalsize{$^{2}$Earth System Division, National Institute for Environmental} \\ 
\normalsize{Studies (NIES), Tsukuba, Japan}\\
\normalsize{$^{3}$Laboratoire des Sciences du Climat et de l’Environnement} \\ 
\normalsize{(LSCE), IPSL, CEA/CNRS/UVSQ, Université Paris-Saclay,} \\ 
\normalsize{Gif-sur-Yvette, France} \\
\normalsize{$^{4}$Research Institute for Global Change/Research Centre for} \\ 
\normalsize{Environmental Modelling and Application, Japan Agency for} \\ 
\normalsize{Marine-Earth Science and Technology (JAMSTEC),} \\ 
\normalsize{Yokohama, Japan} \\
\normalsize{$^\ast$To whom correspondence should be addressed; E-mail: \href{su.xuanming@nies.go.jp}{su.xuanming@nies.go.jp}.}
}

\date{}

\begin{document} 


\baselineskip24pt


\maketitle


\begin{sciabstract}
    The observed global mean surface temperature increase from 1998 to 2012 was slower than that since 1951. The relative contributions of all relevant factors including climate forcers, however, have not been comprehensively analyzed. Using a reduced-complexity climate model and an observationally constrained statistical model, we find that La Ni{\~{n}}a cooling and a descending solar cycle contributed approximately 50\% and 26\% of the total warming slowdown during 1998-2012 compared to 1951-2012. Furthermore, reduced ozone-depleting substances and methane accounted for roughly a quarter of the total warming slowdown, which can be explained by changes in atmospheric concentrations. We identify that human factors played an important role in slowing global warming during 1998-2012, shedding light on the evidence for controlling global warming by reducing greenhouse gas emissions.
\end{sciabstract}


\section*{Introduction}

Observations implied a slow warming (SW) in global mean temperature change (\textsl{$\Delta T$}) from 1998 to 2012 compared to the warming from 1951 to 2012, despite the rapid growth in greenhouse gas (GHG) emissions during the same period \cite{ar5TS}. The warming difference of decadal trends (DDT), that is, the decadal trend of 1998-2012 minus that of 1951-2012, indicated a negative trend. Newer temperature records with updated sea surface temperature (SST) datasets and the infilling of missing data in places such as the Arctic showed greater positive trends, implying that the slowing was not as severe as previously thought \cite{Simmons2017}. For example, the latest Met Office Hadley Centre/Climatic Research Unit global surface temperature anomalies, version 5 (HadCRUT5), used by default in this study, showed an apparently less negative DDT (-0.011~$^{\circ}$C/decade) compared to the older version of HadCRUT4.6 (-0.055~$^{\circ}$C/decade). Such a slowing trend may differ among independent observations (fig. \ref{fig:tem_obsrv}). Nonetheless, there is no doubt that SW does exist \cite{Lean2018}. The Sixth Assessment Report (AR6) of the Intergovernmental Panel on Climate Change (IPCC) referred to this phenomenon as a temporary event \cite{ar6wg1ch3}, primarily due to internal variability \cite{Kosaka2013,Watanabe2014,Brown2015,Dai2015,Steinman2015,Pasini2017} and natural forcings such as volcanic and solar irradiance \cite{Lean2008,Solomon2011,Haywood2014,Ridley2014,Huber2014,Santer2014,Lean2018}. The strengths of internal variability or natural forcings, however, are uncertain \cite{Deser2017,Wang2017,ar5TS,ar6wg1ch3}. Reductions in methane and ozone-depleting substances (ODS) \cite{Estrada2013,Lu2022} or stratospheric water vapour \cite{Solomon2010} may also have contributed, but their respective impacts have not been explicitly calculated. Furthermore, SW cannot be described by a single factor; rather, it necessitates an integrated influence combining multiple components, such as internal variability, forcing changes, ocean heat uptake, and insufficient observational coverage \cite{Huber2014,Schmidt2014,Medhaug2017,Hedemann2017,Power2017}. Quantifications of relative contributions from these individual components are important to interpret the causes of SW.

Physical climate models may insufficiently capture the internal variability or underestimate the response of solar irradiance \cite{Lean2018}, resulting in a higher-than-observed \textsl{$\Delta T$} during the SW period \cite{ar6wg1ch3}. Statistical models, on the other hand, can capture \textsl{$\Delta T$} variations but strongly depend on the chosen predictors and their time lags and usually use the whole anthropogenic influence as an input \cite{Lean2008,Foster2011,Lean2018}, hindering further attributions at the emission level. A systematic and reliable assessment of the individual contributing factors to SW has not been conducted to our knowledge. Here, we combined a reduced-complexity model (RCM) used to quantify the \textsl{$\Delta T$} trends caused by individual climate forcers with a statistical regression model to reconcile the causes with the observed \textsl{$\Delta T$}.


\section*{Results}
\subsection*{Method summary}

\begin{figure}[ht]%
    \centering
    \includegraphics[width=0.9\textwidth]{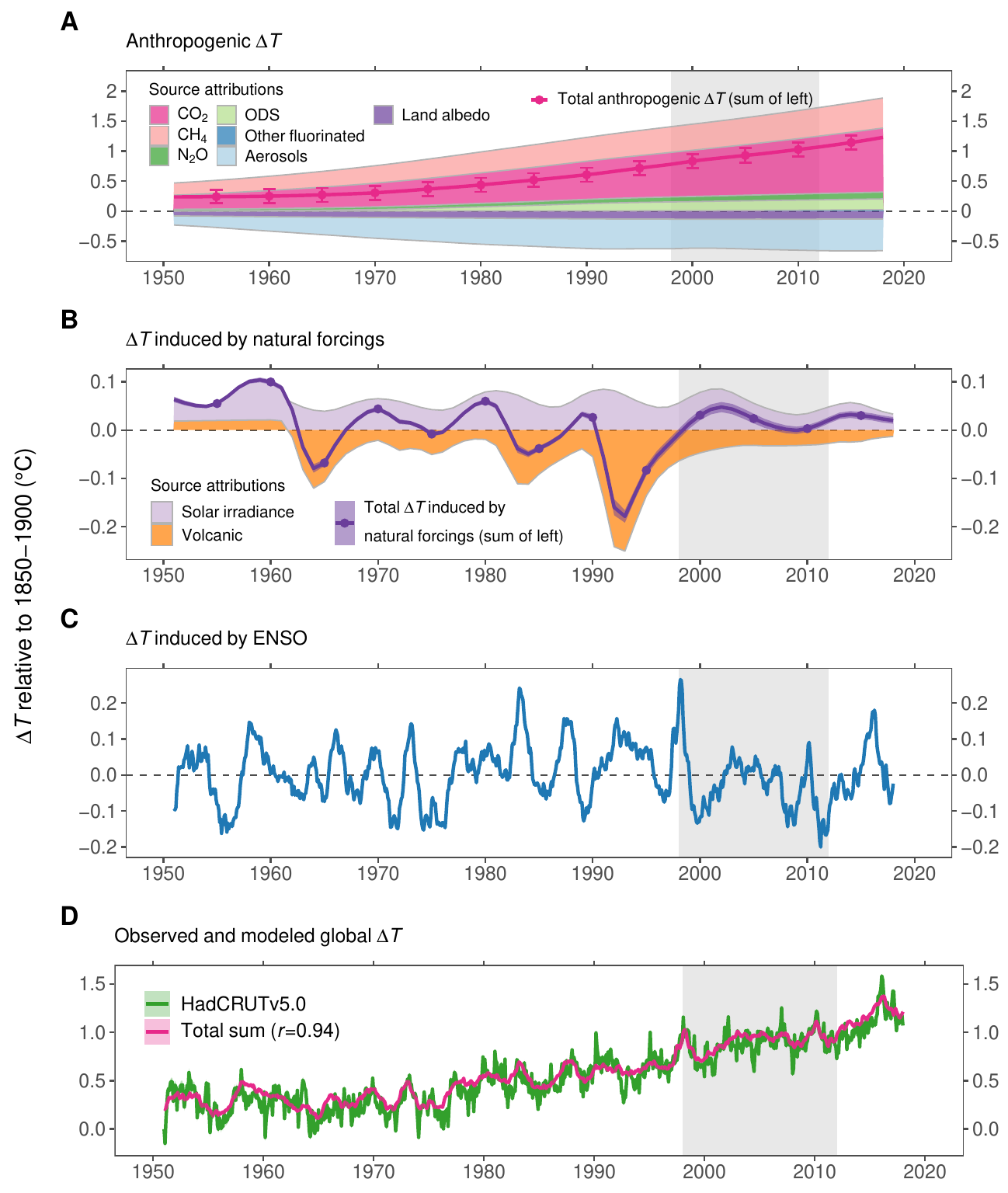}
    \caption{\textbf{Anthropogenic and natural \boldmath{$\Delta T$}.} (\textbf{A}) Anthropogenic \textsl{$\Delta T$}. The thick red solid line and error bars indicate total anthropogenic \textsl{$\Delta T$}. (\textbf{B}) \textsl{$\Delta T$} caused by natural forcings. The thick purple solid line and range represent the total \textsl{$\Delta T$} induced by natural forcings. Color ribbons in (A and B) show \textsl{$\Delta T$} caused by individual factors. (\textbf{C}) \textsl{$\Delta T$} caused by ENSO. (\textbf{D}) HadCRUT5 and the total summed \textsl{$\Delta T$}. Pearson's correlation between the total sum and HadCRUT5 is shown in (D). The error bar in (A) and ranges in (B to D) are the one-sigma produced by the ensemble of statistical regressions. See materials and methods for further information on estimating uncertainty. The SW era from 1998 to 2012 is shaded in light gray.}
    \label{fig:demo}
\end{figure}

To attribute anthropogenic and natural \textsl{$\Delta T$}, first, we applied a normalized marginal approach \cite{Trudinger2005,Li2016,Fu2020} to an RCM, the Simple Climate Model for Optimization version 3.3 (SCM4OPT v3.3) \cite{Su2017, Su2018, Nicholls2021, Su2022}, to quantify the \textsl{$\Delta T$} trends for respective climate forcers (see materials and methods). Second, we used a regression statistical model \cite{Lean2008,Foster2011,Lean2018} to decode the impacts of anthropogenic and natural factors. This was a twofold approach: the regression can correct the biases in the magnitude of the simulated \textsl{$\Delta T$}, while the relative contributions of external forcing factors and internal variability to the observed \textsl{$\Delta T$} are resolved. To avoid overfitting, we tested a group of statistical models with different predictors and chose the one with the lowest Bayesian information criterion (BIC) (table \ref{tab:bic}). The monthly observed \textsl{$\Delta T$} was then modeled as a multiple regression of anthropogenic factors, natural forcings, El Ni{\~{n}}o and Southern Oscillation (ENSO), and residuals allowing for errors in simulations and observations (see materials and methods). As a result, we decomposed the observed \textsl{$\Delta T$} into \textsl{$\Delta T$} caused by each anthropogenic factor, such as CO\textsubscript{2}, CH\textsubscript{4}, N\textsubscript{2}O, ODS, other fluorinated gases, aerosols and pollutants, and land use albedo, as well as each natural factor, such as volcanic eruptions, solar irradiance, and ENSO (Fig. \ref{fig:demo}). There was a high correlation (r = 0.94, 1951-2018) between the modeled \textsl{$\Delta T$} (total sum of individual factors) and the observed \textsl{$\Delta T$}, indicating that most of the perturbation in the observed \textsl{$\Delta T$} was captured in our study.

\pagebreak    
\clearpage

\subsection*{Methane stabilization and ODS mitigation help slow global warming}

\begin{figure}[ht]%
    \centering
    \includegraphics[width=0.9\textwidth]{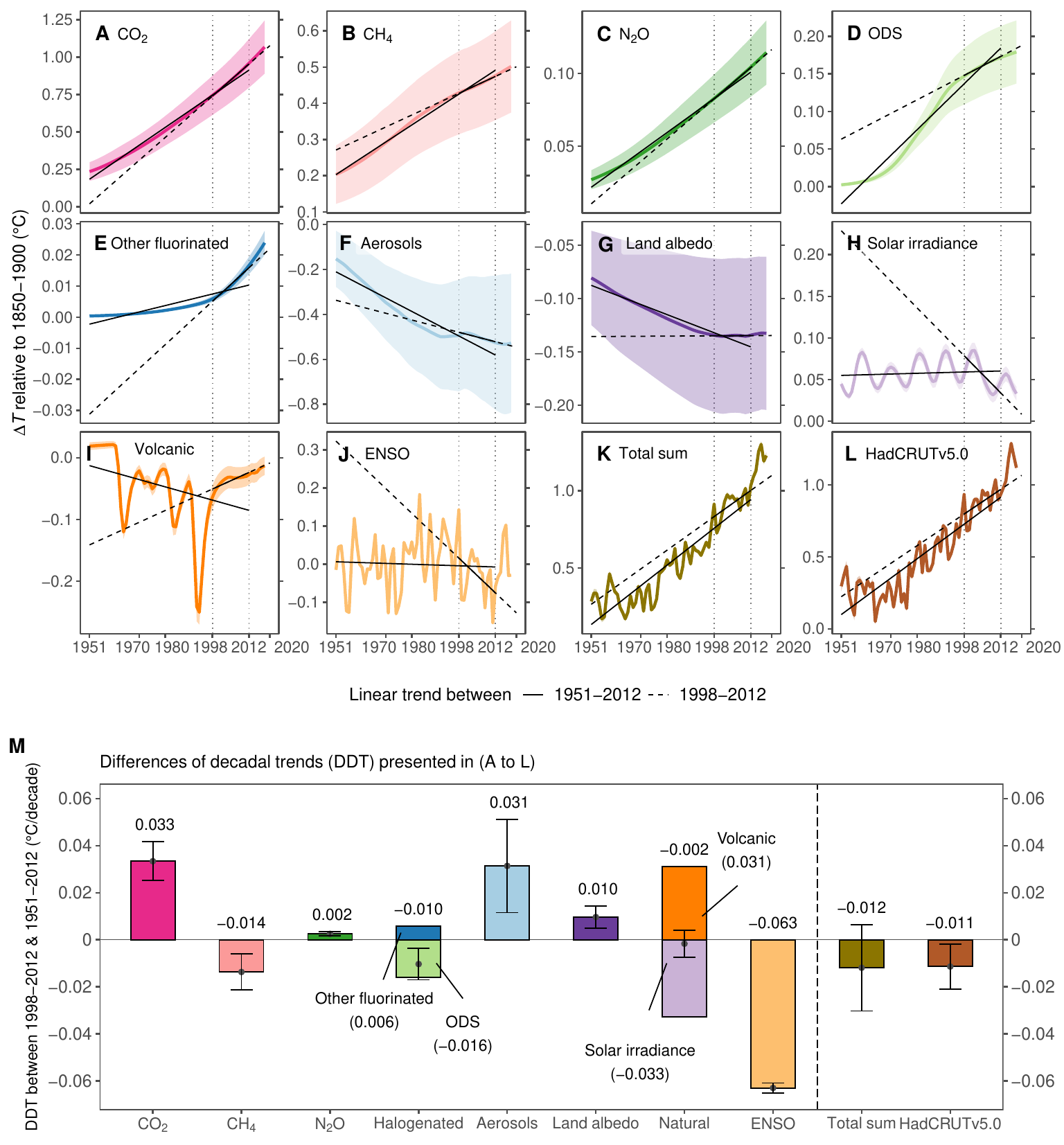}
    \caption{\textbf{Anthropogenic and natural \boldmath{$\Delta T$} by source and their DDT between 1998-2012 and 1951-2012.} (\textbf{A} to \textbf{J}) \textsl{$\Delta T$} and uncertainty ranges caused by individual anthropogenic and natural factors. (\textbf{K} and \textbf{L}) The total sum and HadCRUT5 and their uncertainty ranges. Thin solid and solid with extended dashed lines denote linear trends from 1951 to 2012 and 1998 to 2012, respectively. (\textbf{M}) DDT derived from (A to L). The mean is displayed on top of the bar, and the error bar is the one-sigma produced by the ensemble of statistical regressions.}
    \label{fig:attrib}
\end{figure}

Fig. \ref{fig:attrib}A-L shows \textsl{$\Delta T$} attributions by source, the total sum of all sources, and the observed \textsl{$\Delta T$} HadCRUT5 to aid comparison. Gradients indicated by the thin solid and solid with extended dashed lines represent \textsl{$\Delta T$} linear trends from 1951 to 2012 and from 1998 to 2012, respectively. Accordingly, to the left of the junction, if the dashed line is above the solid line, it implies that the growth rate is slowing, and vice versa. Methane, ODS, solar irradiance, and ENSO negatively contributed to \textsl{$\Delta T$} during 1998-2012 compared to 1951-2012. The emission attributions of \textsl{$\Delta T$} agree mostly with the AR6 forcing-based estimate \cite{ar6wg1spm} (fig. \ref{fig:compare2}), except for halogenated \textsl{$\Delta T$}, which is approximately twice as high as AR6. This is because our method generated a smaller negative ozone effective radiative forcing (\textsl{$\Delta F$}) (-0.05 Wm\textsuperscript{-2} out of total ozone forcing 0.26 Wm\textsuperscript{-2}, 1850-2019), which was induced by halogenated gases (cf. -0.16 Wm\textsuperscript{-2} out of 0.47 Wm\textsuperscript{-2}, 1750-2019 \cite{Thornhill2021}, used in AR6) (see fig. \ref{fig:o3} for the attribution of ozone effective \textsl{$\Delta F$}). Consequently, a smaller \textsl{$\Delta T$} was mitigated by ozone, and halogenated \textsl{$\Delta T$} was positively larger in our results.

Based on the attributed \textsl{$\Delta T$}, we computed the decadal trend (fig. \ref{fig:hist_hd50}) and DDT (Fig. \ref{fig:attrib}M) for each source. Note that DDT was also affected by the referred historical period; selecting a smaller period of 1960-2012 would highlight such a slowdown more clearly because \textsl{$\Delta T$} increased more rapidly throughout 1960-2012 (fig. \ref{fig:tem_obsrv}). However, we examined 1951-2012 for a general case. 

ENSO contributed the most to the SW, with a DDT of -0.063$\pm$0.002~$^{\circ}$C/decade, followed by solar irradiance (-0.033$\pm$0.002~$^{\circ}$C/decade). Anthropogenic ODS (-0.016$\pm$0.007~$^{\circ}$C/decade) and methane (-0.014$\pm$0.008~$^{\circ}$C/decade) also contributed to the SW. Specifically, ENSO and solar variations diminished warming directly with a negative decadal trend, while ODS and methane emissions simply exhibited a smaller decadal trend during 1998-2012 compared to that of 1951-2012. Other sources, including CO\textsubscript{2}, N\textsubscript{2}O, other fluorinated gases, aerosols and pollutants, land albedo, and volcanic activity, showed positive DDT, meaning that they increased faster between 1998 and 2012 than they did between 1951 and 2012. It is worth noting that the \textsl{$\Delta T$} of total GHG emissions increased steadily (DDT 0.012$\pm$0.015~$^{\circ}$C/decade), masking the slowdown of ODS and methane \textsl{$\Delta T$} (fig. \ref{fig:trd_ghg}), which is likely why previous research missed them. In addition, aerosols and pollutants and land albedo (see fig. \ref{fig:rfc_lcc} for the forcing trends) contributed to negative warming and declined over 1951-2012. However, during the SW period, their downward tendencies slowed and contributed to a positive DDT (see fig. \ref{fig:dif_regre_hd50}).

\pagebreak    
\clearpage

\begin{figure}[ht]%
    \centering 
    \includegraphics[width=0.9\textwidth]{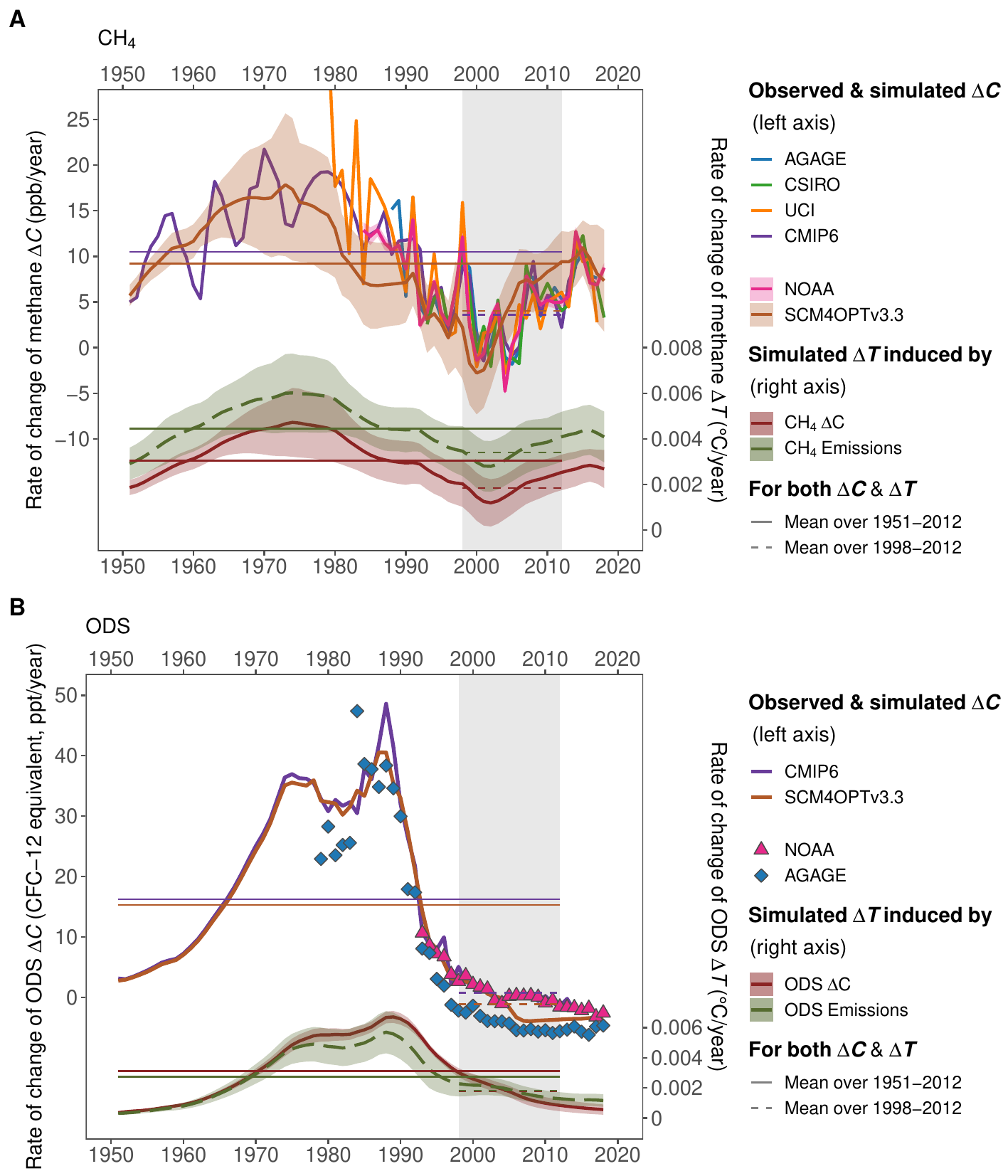}
    \caption{\textbf{Rate of change of methane and ODS \boldmath{$\Delta C$} and the induced \boldmath{$\Delta T$}}. (\textbf{A}) Methane and (\textbf{B}) ODS. The SCM4OPT v3.3 uncertainties in (A and B) are the one-sigma of the Monte Carlo simulation with $n = 1000$. \textsl{$\Delta T$} uncertainty ranges in (A and B) are the one-sigma produced by the ensemble of statistical regressions. In (B), the ODS components for each source differ (table \ref{tab:comp}). Uncertainty is not considered in ODS \textsl{$\Delta C$} because it is an equivalent value combining a collection of compounds of ODS \cite{Meinshausen2017}. CMIP6 \cite{Meinshausen2017} and SCM4OPT v3.3 contain data for the entire evaluation period. Mean values from 1951 to 2012 and 1998 to 2012 are illustrated by thin horizontal solid and dashed lines, respectively. The SW era from 1998 to 2012 is shaded in light gray.}
    \label{fig:cause}
\end{figure}

The slowdown of methane \textsl{$\Delta T$} can be explained by the associated atmospheric concentrations (\textsl{$\Delta C$}). Growths in methane \textsl{$\Delta C$} is reported to have slowed from the 1980s and to have stabilized between 1999 and 2006 \cite{Nisbet2014,Nisbet2016,Rigby2008,Dlugokencky2009,Montzka2011,Meinshausen2017}, owing in part to reductions in emissions from oil-gas exploitation and enteric fermentation in Europe and Russia \cite{Naveen2021}. As shown in Fig. \ref{fig:cause}A, the DDT of methane \textsl{$\Delta C$} simulated by SCM4OPT v3.3 presented a negative value of -52.4 ppb/decade, indicating that growths in methane \textsl{$\Delta C$} slowed significantly during the SW period, which partly explains the slowdown of methane \textsl{$\Delta T$} at the same time. Our results do not reflect methane emissions from natural sources, which accounted for approximately 39\% of overall methane emissions in the recent decade (ref. \cite{Saunois2020}, top-down estimate for 2008–2017). Furthermore, methane emissions from ultraemitters in the oil and gas industry, accounting for up to 12\% of global methane emissions from oil and gas production and transmission \cite{Lauvaux2022}, were not included in our calculations. Hence, the actual methane \textsl{$\Delta C$} (for instance, DDT of CMIP6 \textsl{$\Delta C$} -88.6 ppb/decade) may have contributed a more negative DDT in methane \textsl{$\Delta T$}. 

ODS manifested consistent DDT between the simulated \textsl{$\Delta C$} (-224.9 CFC-12 equivalent ppt/decade) and the observed \textsl{$\Delta C$} (CMIP6 -212.5 CFC-12 equivalent ppt/decade) (Fig. \ref{fig:cause}B) because they are well documented \cite{Montzka2011ods,Montzka2011,Meinshausen2017}; hence, \textsl{$\Delta C$} can be well captured by the model. ODS have been effectively mitigated under the Montreal Protocol since the late 1980s \cite{ipccar4ch2}, resulting in a stabilized mixing ratio in the atmosphere. Such a tendency can also be used to interpret the negative DDT of ODS \textsl{$\Delta T$}. The \textsl{$\Delta C$} of other GHGs including CO\textsubscript{2}, N\textsubscript{2}O and other fluorinated gases, however, continued to rise, accelerating \textsl{$\Delta T$} rather than slowing it during the SW era (fig. \ref{fig:conc_trd}).

The emission-based attribution takes chemical-physical changes in the atmosphere into account. Methane \textsl{$\Delta T$}, for example, includes \textsl{$\Delta T$} produced by atmospheric methane \textsl{$\Delta C$}, stratospheric water vapour from methane oxidation, the feedback on tropospheric ozone and on sinks of halogenated gases, and CO\textsubscript{2} from oxidized methane. Concerning ODS \textsl{$\Delta T$}, their contributions contain \textsl{$\Delta T$} caused by ODS \textsl{$\Delta C$} and its feedback on stratospheric ozone and on methane sinks (Fig. \ref{fig:cause}B). A minor difference might exist between the \textsl{$\Delta T$} resulting from \textsl{$\Delta C$} and the associated emissions, but their major trends are consistent (Fig. \ref{fig:cause}, fig. \ref{fig:conc_trd}). Therefore, the slowdown of ODS and methane \textsl{$\Delta T$} during the SW can be robustly explained by their slowing trends in atmospheric \textsl{$\Delta C$}.  It is important to note that the slowing trends for both \textsl{$\Delta C$} and \textsl{$\Delta T$} induced by ODS and methane occurred earlier than 1998-2012, which also contributed to the SW.

\pagebreak    
\clearpage

\subsection*{Contributions of natural forcing factors and ENSO}

\begin{figure}[ht]%
    \centering
    \includegraphics[width=0.9\textwidth]{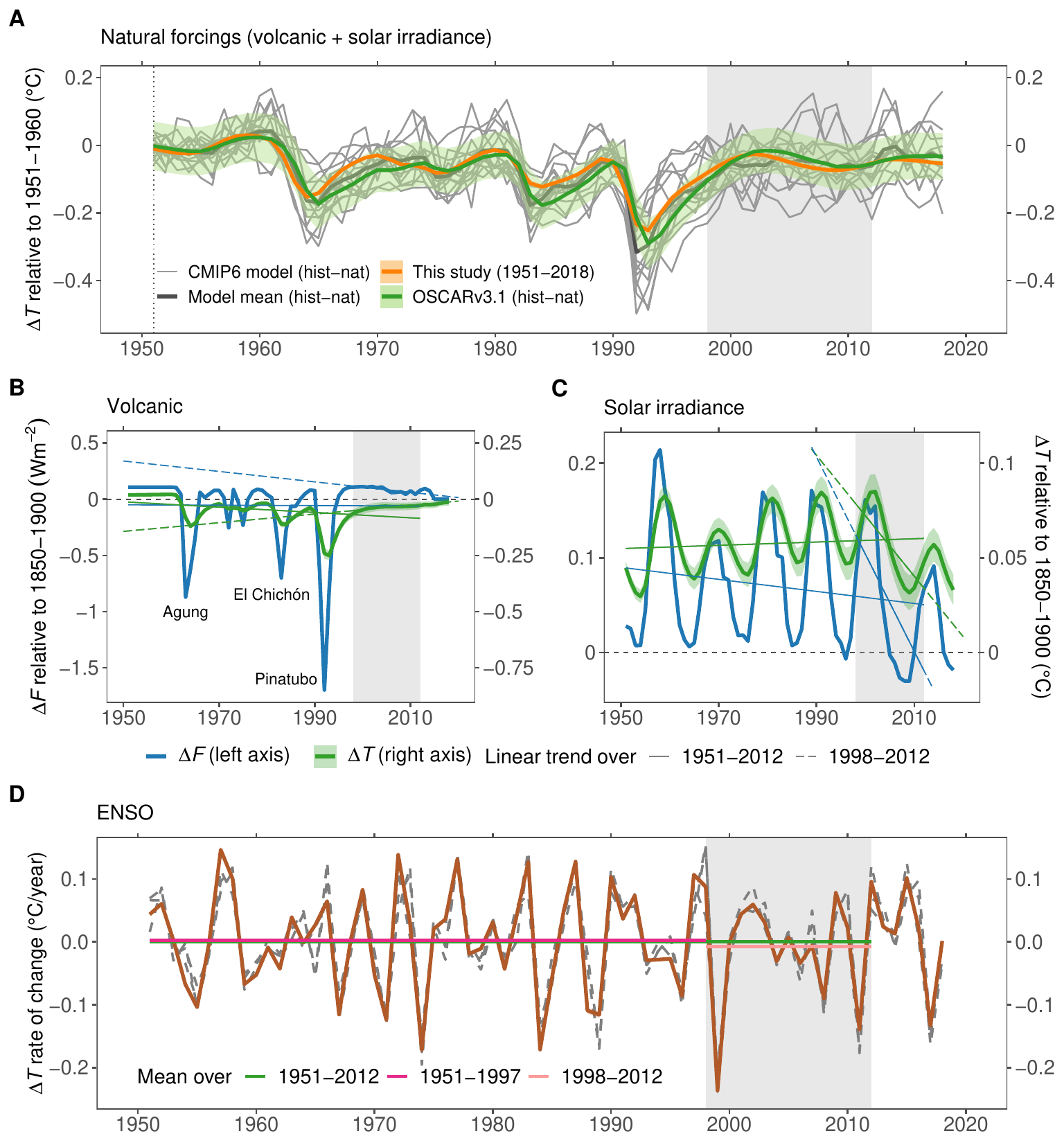}
    \caption{\textbf{\boldmath{$\Delta T$} due to natural forces.} (\textbf{A}) Natural forcing \textsl{$\Delta T$} from OSCAR v3.1 \cite{Gasser2020} and CMIP6 Earth System Models (ESMs) \cite{Gillett2016} (ensemble mean for each ESM, table \ref{tab:cmip6}), as well as this study. OSCAR v3.1 uncertainty denotes the one-sigma obtained from Monte Carlo simulations with n = 1000. (\textbf{B} and \textbf{C}) \textsl{$\Delta F$} and \textsl{$\Delta T$} for volcanic eruptions and solar irradiance. In (B and C), thin solid and solid with extended dashed lines represent linear trends for the years 1951-2012 and 1998-2012, respectively. Three large volcanic eruptions are indicated in (B). (\textbf{D}) \textsl{$\Delta T$} rate of change caused by ENSO (see fig. \ref{fig:vari} for the ENSO \textsl{$\Delta T$} trend). Gray dashed lines in (D) depict \textsl{$\Delta T$} rates of changes estimated by using other ENSO indices shown in figs. \ref{fig:vari_ncep}-\ref{fig:vari_best}. Thick horizontal lines represent the mean values for the specified periods. The \textsl{$\Delta T$} ranges of this study in (A to D) are the one-sigma produced by the ensemble of statistical regressions. The SW era from 1998 to 2012 is shaded in light gray.} 
    \label{fig:nat}
\end{figure}

As shown in Fig. \ref{fig:nat}A, our natural forcing \textsl{$\Delta T$} is in line with results from previous RCMs \cite{Gasser2020} as well as more complex Earth system models (ESMs) \cite{Gillett2016} (for GHGs and aerosols and pollutants, see figs. \ref{fig:trd_ghg}\&\ref{fig:trd_aer}). Fig. \ref{fig:nat}B depicts volcanic \textsl{$\Delta F$} and \textsl{$\Delta T$}. Although volcanic \textsl{$\Delta F$} had a negative DDT (-0.045 Wm\textsuperscript{-2}/decade), recovery from the cooling caused by large volcanic activities in the 20\textsuperscript{th} century caused a long-term negative trend (-0.012~$^{\circ}$C/decade, 1951-2012), whereas weak volcanic activity in the 1998-2012 period caused a small positive trend (0.019~$^{\circ}$C/decade, 1998-2012), resulting in a positive DDT. In the case of solar irradiance (Fig. \ref{fig:nat}C), the decadal trends for both \textsl{$\Delta F$} and \textsl{$\Delta T$} became slower during 1998-2012 compared to 1951-2012. Thus, we can pin at least some of the cooling on the descending solar cycle. Regarding volcanic aerosols, in contrast to earlier studies \cite{Solomon2011,Huber2014,Ridley2014}, which could balance out up to 30\% of anthropogenic \textsl{$\Delta T$} \cite{Lean2018}, our results suggest that volcanic forcing is not responsible for the SW. Our result also implies a relatively large contribution from the descending solar cycle, similar to Lean (2018) \cite{Lean2018}, which mitigated approximately 36\% of anthropogenic \textsl{$\Delta T$} from 2001 to 2011. However, Lean (2018) also showed a small 8\% cut in anthropogenic \textsl{$\Delta T$} from volcanic aerosols at the same time, possibly due to the use of stratospheric aerosol optical depth (AOD) to estimate volcanic \textsl{$\Delta T$} in her study, which is a proxy of volcanic aerosol \textsl{$\Delta F$}. The volcanic \textsl{$\Delta F$} actually presented a relatively small negative decadal trend (-0.046 Wm\textsuperscript{-2}/decade, blue thin dashed line in Fig. \ref{fig:nat}B) from 1998-2012, which could produce the small volcanic cooling effect reported by Lean (2018).

ENSO was a major contributor to the SW, which was mainly led by strong La Ni{\~{n}}a cooling immediately after El Ni{\~{n}}o warming prior to 1998 \cite{Kosaka2013,ar6wg1ch3} (Fig. \ref{fig:nat}D). Because of its seasonality, ENSO contributed to SW in the form of an annual pulse signal (figs. \ref{fig:vari}-\ref{fig:vari_best}), whereas the descending solar cycle, ODS, and methane had a relatively continuous and stable influence on SW. Our result is consistent with that of Lean (2018), which signifies the largest contribution from ENSO, with a cooling decadal trend of -0.086~$^{\circ}$C/decade from 2001-2011 (cf. -0.065~$^{\circ}$C/decade from 1998-2012 in this study, see fig. \ref{fig:hist_hd50}).  

\pagebreak    
\clearpage

\section*{Discussion and conclusions}

The total sum of the DDT of individual factors converged on the observation (Fig. \ref{fig:attrib}M), explaining almost all of the \textsl{$\Delta T$} slowdown during the SW period. The use of different independent observations (figs. \ref{fig:attrib_berk}-\ref{fig:attrib_glb}) or ENSO indices (figs. \ref{fig:attrib_hd50_ncep}-\ref{fig:attrib_hd50_best}) gave similar results, implying that our analysis is quantitatively robust. The uncertainties may result from the autocorrelation of chosen predictors and cross-correlation among the predictors, as well as observational errors or inaccurate depictions of the real processes \cite{Lean2018}. The correlation between simulated \textsl{$\Delta T$} and observed \textsl{$\Delta T$} here is higher than Lean's (2018) statistical method (r= 0.91 for space-based model during 1979–2017), most likely due to the full consideration of all available factors, which were derived from a physical RCM. This is also due to the use of the latest temperature record, while the older version may overmeasure the SW, resulting in larger residuals and lower goodness-of-fit (lower R-squared values) in the regression (fig. \ref{fig:attrib_hd46}, table \ref{tab:coef}). Therefore, our causal explanation of the SW is sound.

Combining a physical RCM and an observationally constrained statistical model, we comprehensively evaluated the relative contributions to the SW from 1998-2012. Using DDT, we captured not only the direct cooling contributions but also the slowing trends of \textsl{$\Delta T$} in the SW caused by anthropogenic and natural factors. Our results show that carbon dioxide, aerosols and pollutants and erupting volcanoes significantly accelerated global \textsl{$\Delta T$} during the SW period compared to 1951-2012, whereas La Ni{\~{n}}a cooling and a descending solar cycle strongly offset the increased \textsl{$\Delta T$}, accounting for approximately 50\% and 26\% of the total warming slowdown during 1998-2012 compared to 1951-2012. In particular, we identified that the reduced anthropogenic ODS and methane emissions also contributed approximately 13\% and 11\% of the total warming slowdown, which can be explained by the recorded atmospheric \textsl{$\Delta C$}. Various factors superimposed on the timeline slowed \textsl{$\Delta T$} between 1998 and 2012, while the beginning and end of a single source did not necessarily match the SW period. The contribution of anthropogenic ODS and methane was comparable in scale to La Ni{\~{n}}a cooling and the descending solar cycle, as well as the overall downward trend exhibited by the temperature records (Fig. \ref{fig:attrib}M), all of which can be detected with instruments, either directly or indirectly. Thus, our findings provide practical evidence for preventing global warming by reducing GHG emissions.



\bibliography{SlowWarm}
\bibliographystyle{Science}

\section*{Acknowledgments}
The computational resources were provided by the National Institute for Environmental Studies (NIES), Japan. \textbf{Funding:} This work was supported by the Decarbonized and Sustainable Society Research Program at the National Institute for Environmental Studies, Japan, and the Program for the advanced studies of climate change projection (SENTAN, Grant Number JPMXD0722681344) from the Ministry of Education, Culture, Sports, Science and Technology (MEXT), Japan. K. Tanaka benefited from state assistance managed by the National Research Agency in France under the Programme d’Investissements d’Avenir under the reference ANR-19-MPGA-0008. \textbf{Author contributions:} X.S. designed the study, performed the simulations, produced all figures, and led the writing with contributions from all authors. H.S. provided insights to create the statistical model. \textbf{Competing interests:} The authors declare no competing interests. \textbf{Data and materials availability:} The data used in this analysis can be accessed online (last access for all, 22 May 2023): Coupled Model Intercomparison Project 6 (CMIP6) (\url{https://esgf-node.llnl.gov/projects/cmip6/}); Community Emissions Data System (CEDS) (\url{https://esgf-node.llnl.gov/search/input4mips/}); Emission Database for Global Atmospheric Research (EDGAR)  v7.0 (Global Greenhouse Gas Emissions) (\url{https://edgar.jrc.ec.europa.eu/dataset_ghg70}); EDGAR v6.1 (Global Air Pollutant Emissions) (\url{https://edgar.jrc.ec.europa.eu/dataset_ap61}); Coupled Model Intercomparison Project Phase 6 (CMIP6) emissions from the Integrated Assessment Modeling Consortium (IAMC) (\url{https://esgf-node.llnl.gov/search/input4mips/}); The Atmospheric Chemistry and Climate Model Intercomparison Project (ACCMIP) \cite{Lamarque2013} (\url{https://tntcat.iiasa.ac.at/RcpDb}); Carbon emissions from land use and land-cover change \cite{Houghton2012,Hansis2015} (\url{http://www.globalcarbonatlas.org/en/CO2-emissions}); Historical landcover change and wood harvest CO\textsubscript{2} emissions from the Max Planck Institute for Meteorology (MPI-M) (\url{https://www.mpimet.mpg.de/en/science/the-land-in-the-earth-system/working-groups/climate-biogeosphere-interaction/landuse-change-emission-data/}); Land-use change emissions from ref \cite{Smith2013} (\url{https://cdiac.ess-dive.lbl.gov/ftp/Smith_Rothwell_Land-Use_Change_Emissions/}); Land-Use Harmonization (LUH2) (\url{https://luh.umd.edu/}); Historical greenhouse gas concentrations for climate modelling (CMIP6) (\url{https://doi.org/10.5194/gmd-10-2057-2017}); National Oceanic \& Atmospheric Administration Carbon Cycle Greenhouse Gases (atmospheric concentrations of CO\textsubscript{2}, CH\textsubscript{4} and N\textsubscript{2}O) (\url{https://gml.noaa.gov/ccgg/}); The NOAA Ozone Depleting Gas Index (\url{https://gml.noaa.gov/odgi/}); Advanced Global Atmospheric Gases Experiment (AGAGE) (\url{https://agage.mit.edu/data/agage-data}); The Global Carbon Project (GCP) \cite{Friedlingstein2020} (https://www.globalcarbonproject.org); Global Methane Budget (\url{https://www.globalcarbonproject.org/methanebudget/}); Hadley Centre/Climatic Research Unit Temperature (HadCRUT) v5.0 (\url{https://www.metoffice.gov.uk/hadobs/hadcrut5/}); Berkeley Earth (Global Temperature Data) (\url{https://berkeleyearth.org/data/}); NOAAGlobalTemp v5.1 (\url{https://www.ncei.noaa.gov/products/land-based-station/noaa-global-temp}); GISS Surface Temperature Analysis (GISTEMP v4) (\url{https://data.giss.nasa.gov/gistemp/}); NOAA MEI \cite{Wolter2011} (\url{https://psl.noaa.gov/enso/mei.old/mei.html}); MEI from NCEP-NCAR (\url{https://www.webberweather.com/multivariate-enso-index.html}); Cold Tongue Index (CTI) (\url{https://github.com/ToddMitchellGH/Cold-Tongue-Index}); ``BEST" ENSO Index \cite{Smith2000} (\url{https://psl.noaa.gov/people/cathy.smith/best/}). The code for reproducing the study will be deposited in the permanent repository Zenodo.


\pagebreak    
\clearpage

\setcounter{page}{1}

\setcounter{figure}{0}
\makeatletter
\renewcommand{\thefigure}{S\@arabic\c@figure}
\makeatother

\setcounter{table}{0}
\makeatletter
\renewcommand{\thetable}{S\@arabic\c@table}
\makeatother

\setcounter{equation}{0}
\makeatletter
\renewcommand{\theequation}{S\@arabic\c@equation}
\makeatother

\section*{Supplementary Materials for}
\begin{center}
\Huge{Anthropogenic contributions to slow warming over 1998-2012}
\end{center}

\begin{center}
\Large{Xuanming Su {\it et al.}}
\end{center}

\begin{center}
Corresponding author: Xuanming Su, \href{su.xuanming@nies.go.jp}{su.xuanming@nies.go.jp}
\end{center}

\begin{figure}[ht]%
    \centering
    \includegraphics[width=0.9\textwidth]{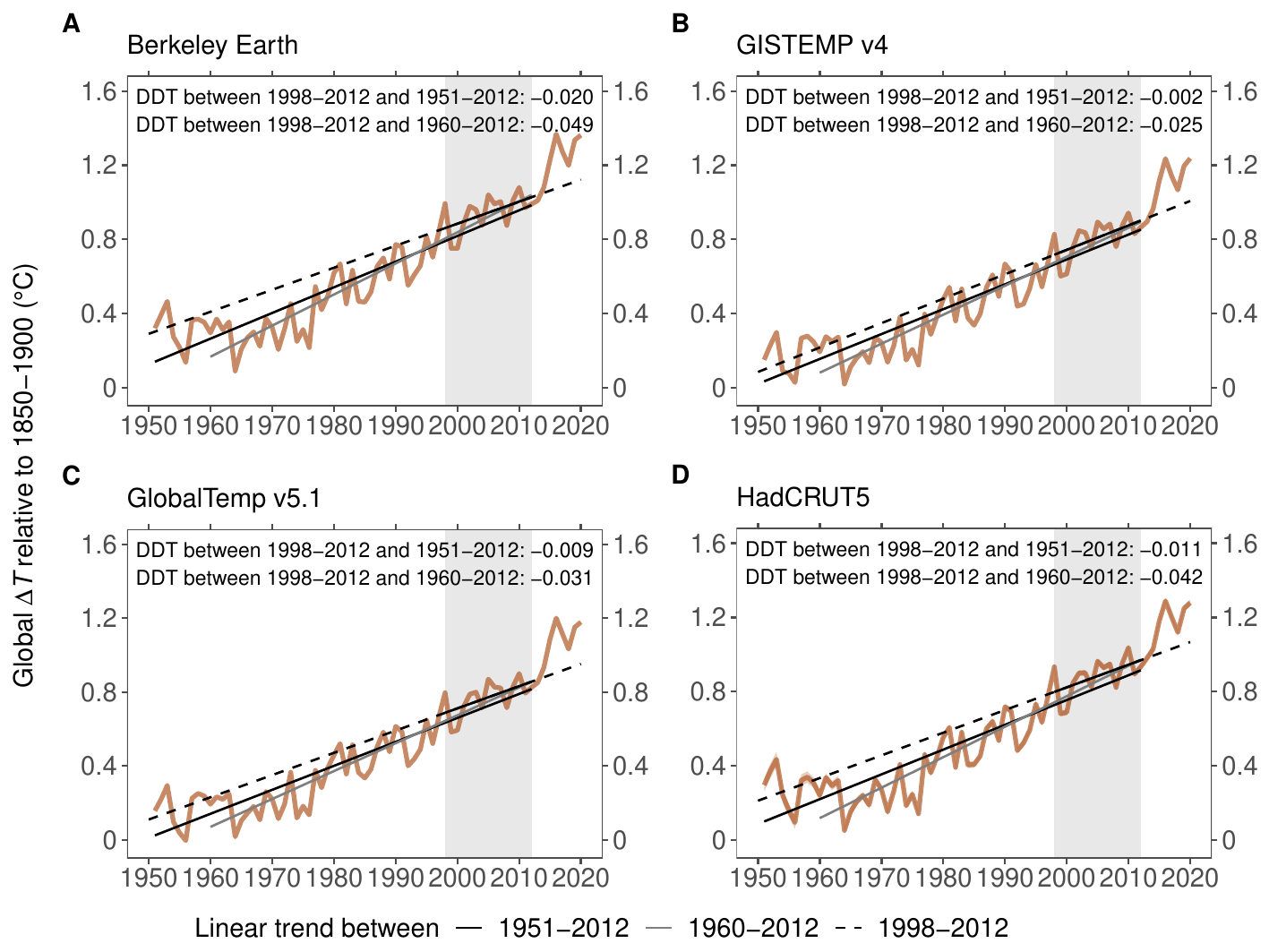}
    \caption{\textbf{Observed \boldmath{$\Delta T$}.} Linear trends across the selected periods are represented by solid and solid with extended dashed lines. \textsl{$\Delta T$} of GISTEMPv4 is relative to 1880-1900 since there are no data available before 1880. Differences of decadal trends (DDT) are shown in the upper part of the panel for each temperature record. The SW era from 1998 to 2012 is shaded in light gray.}
    \label{fig:tem_obsrv}
\end{figure}

\begin{figure}[ht]%
    \centering
    \includegraphics[width=0.75\textwidth]{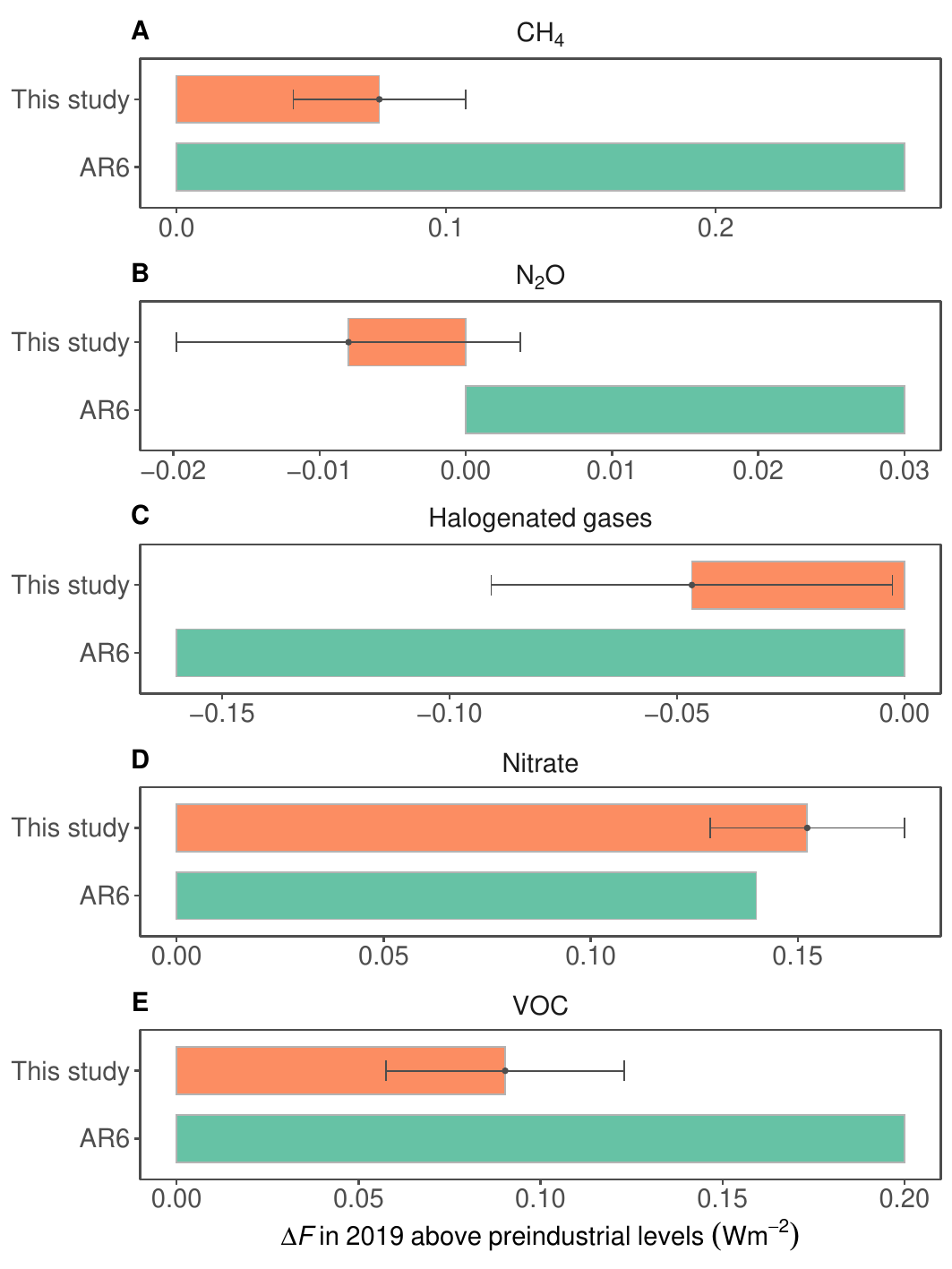}
    \caption{\textbf{Attribution of ozone effective \boldmath{$\Delta F$} in this study compared to AR6.} The effective \textsl{$\Delta F$} for AR6 was extracted from \protect\url{https://github.com/sarambl/AR6\_CH6\_RCMIPFIGS}. This study's evaluation period is 1850-2019, while AR6's evaluation period is 1750-2019. The effective \textsl{$\Delta F$} for this study includes effects from both stratospheric and tropospheric ozone. This study's uncertainty ranges represent the one-sigma of the Monte Carlo simulation with n = 1000; however, the AR6 uncertainty ranges are not included due to data availability. }
    \label{fig:o3}
\end{figure}

\begin{figure}[ht]%
    \centering
    \includegraphics[width=0.9\textwidth]{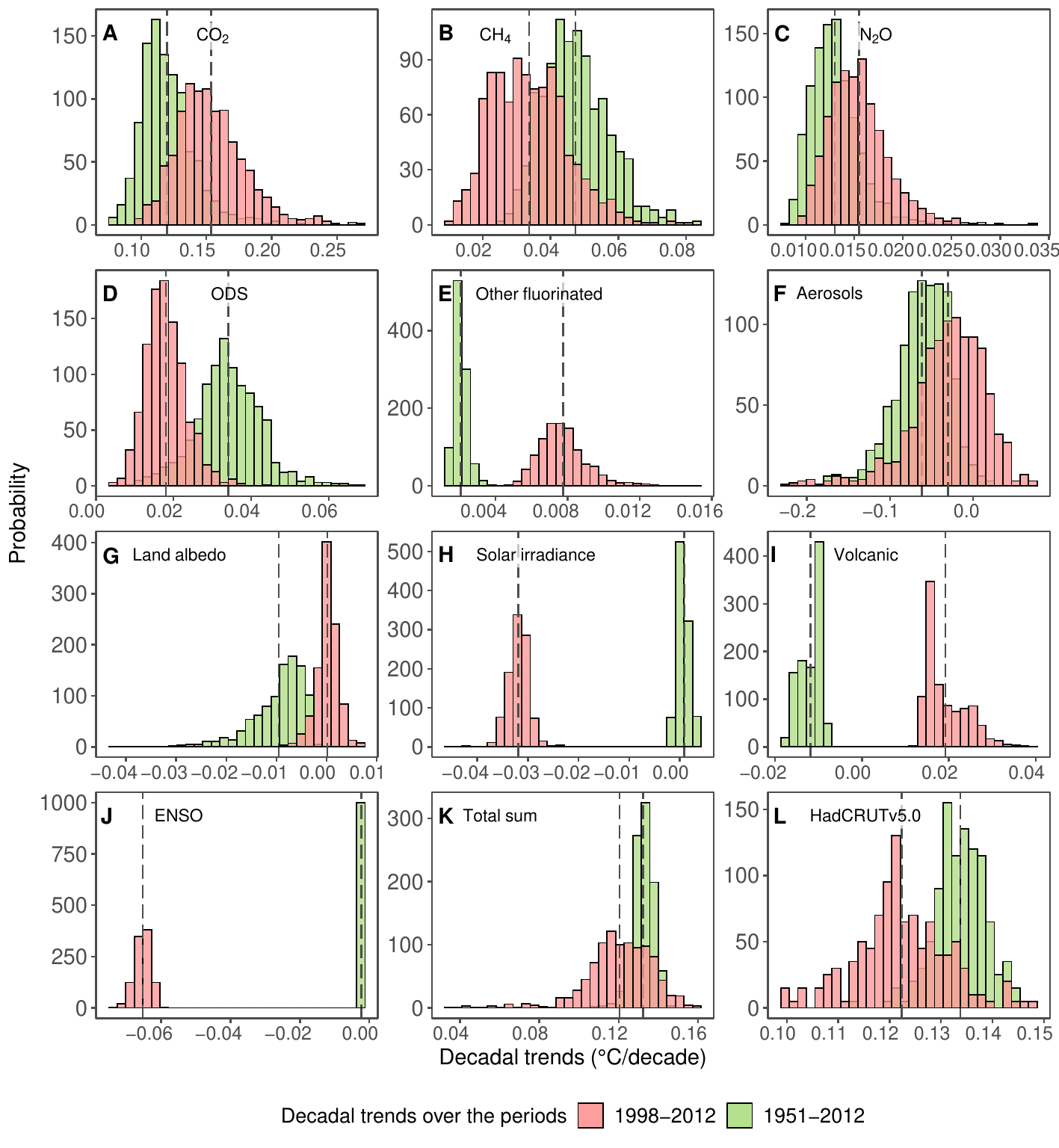}
    \caption{\textbf{Decadal \boldmath{$\Delta T$} trends.} The decadal \textsl{$\Delta T$} trends are shown for the periods of 1998-2012 (red) and 1951-2012 (green). The histogram in (A to K) represents the distribution produced by the ensemble of statistical regressions, while the histogram in (L) shows the 200 realizations of the HadCRUT5 ensemble. The ensemble means are represented by vertical dashed lines over the selected periods. The ENSO is derived from HadCRUT5.}
    \label{fig:hist_hd50}
\end{figure}

\begin{figure}[ht]%
    \centering
    \includegraphics[width=0.9\textwidth]{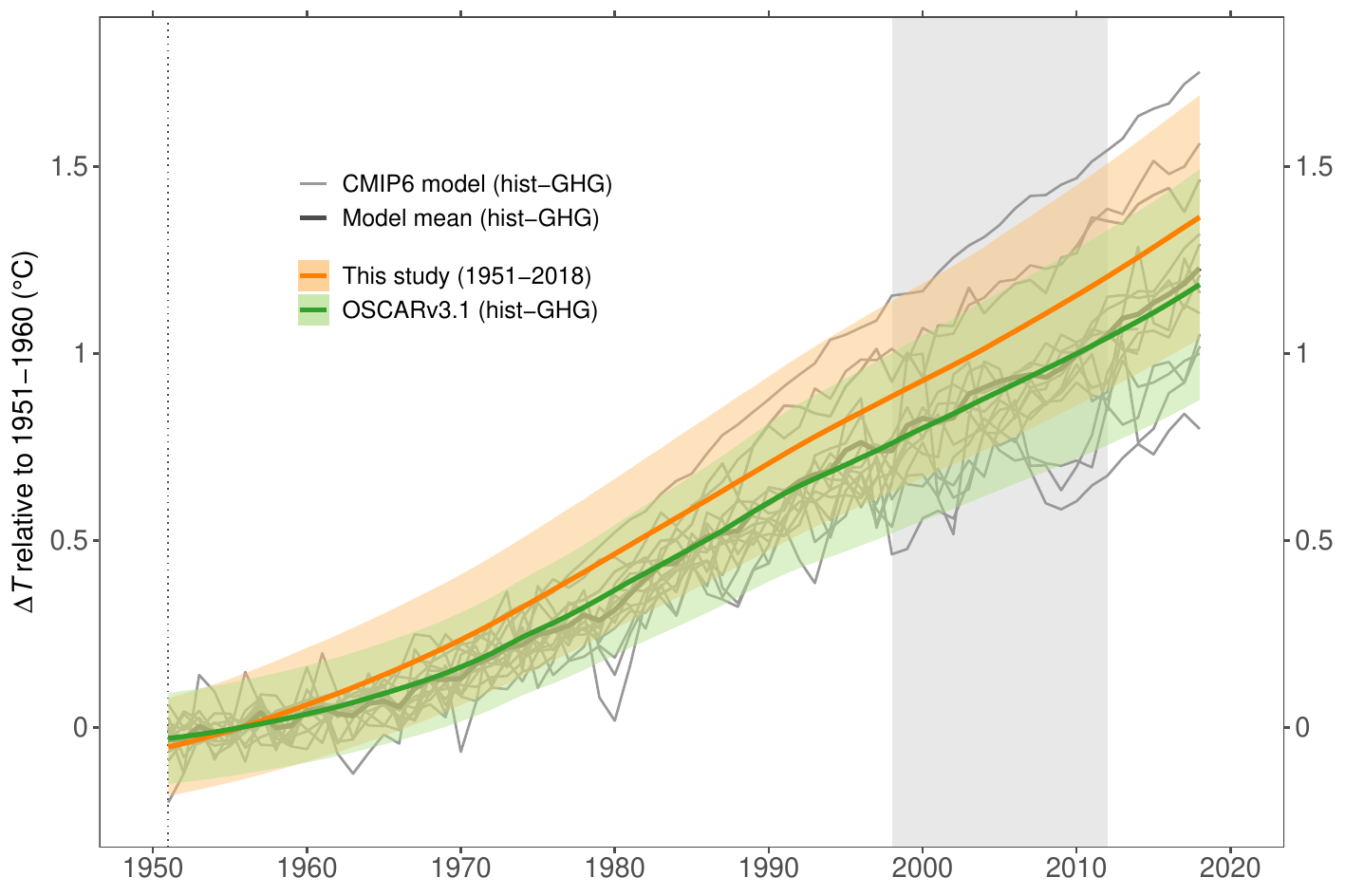}
    \caption{\textbf{\boldmath{$\Delta T$} caused by GHG emissions.} See Fig. \ref{fig:nat} for the uncertainties of this study and OSCAR v3.1 \cite{Gasser2020}. The SW era from 1998 to 2012 is shaded in light gray. Note that the hist-GHG simulation does not account for stratospheric and tropospheric ozone. See table \ref{tab:cmip6} for the investigated CMIP6 ESMs \cite{Gillett2016}.}
    \label{fig:trd_ghg}
\end{figure}

\begin{figure}[ht]%
    \centering
    \includegraphics[width=0.9\textwidth]{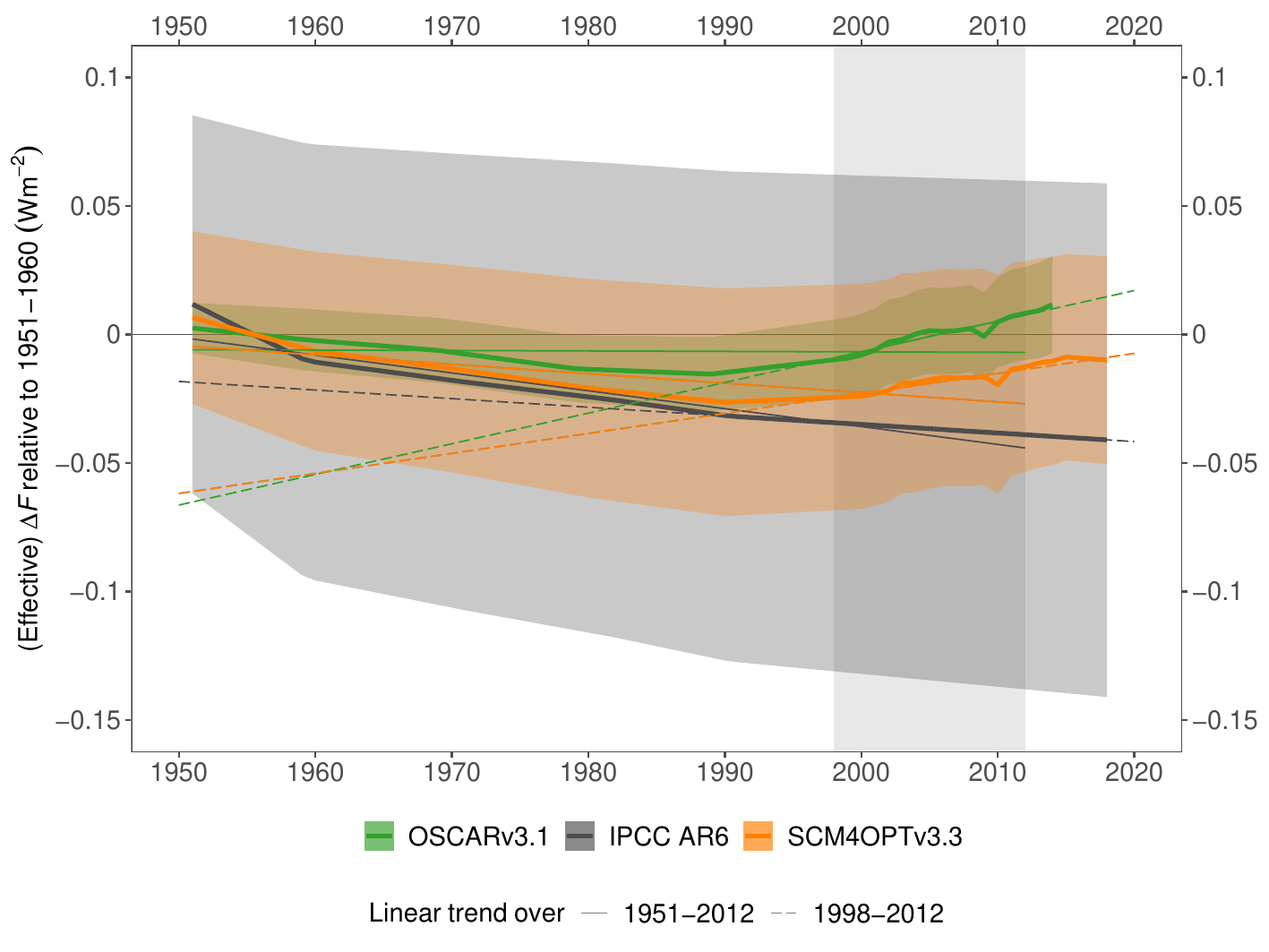}
    \caption{\textbf{(Effective) \boldmath{$\Delta F$} by anthropogenic surface albedo change.} IPCC AR6 \cite{ar6wg1anniii} and SCM4OPT v3.3 are effective \textsl{$\Delta F$}, whereas OSCARv3.1 \cite{Gasser2020} is \textsl{$\Delta F$} due to data availability.}
    \label{fig:rfc_lcc}
\end{figure}

\begin{figure}[ht]%
    \centering
    \includegraphics[width=0.9\textwidth]{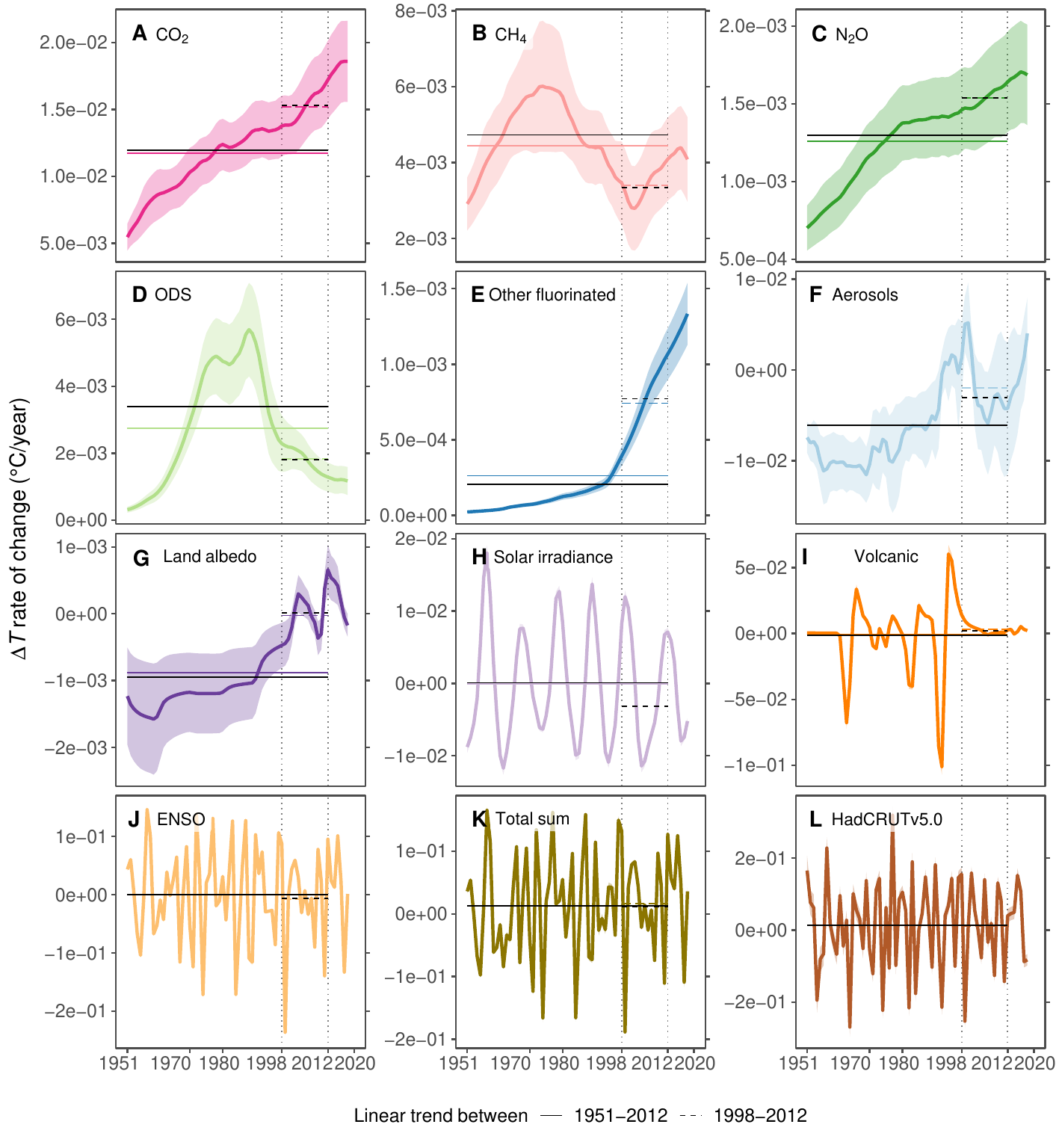}
    \caption{\textbf{\boldmath{$\Delta T$} rate of change.} Color horizontal solid and dashed lines represent mean values from 1951 to 2012 and from 1998 to 2012, respectively. Black horizontal solid and dashed lines represent the linear temperature trends from 1951 to 2012 and 1998 to 2012, respectively (Fig. \ref{fig:attrib} (A to L)). See Fig. \ref{fig:attrib} for the uncertainty description. HadCRUT5 was used to calculate the ENSO.}
    \label{fig:dif_regre_hd50}
\end{figure}

\begin{figure}[ht]%
    \centering
    \includegraphics[width=0.9\textwidth]{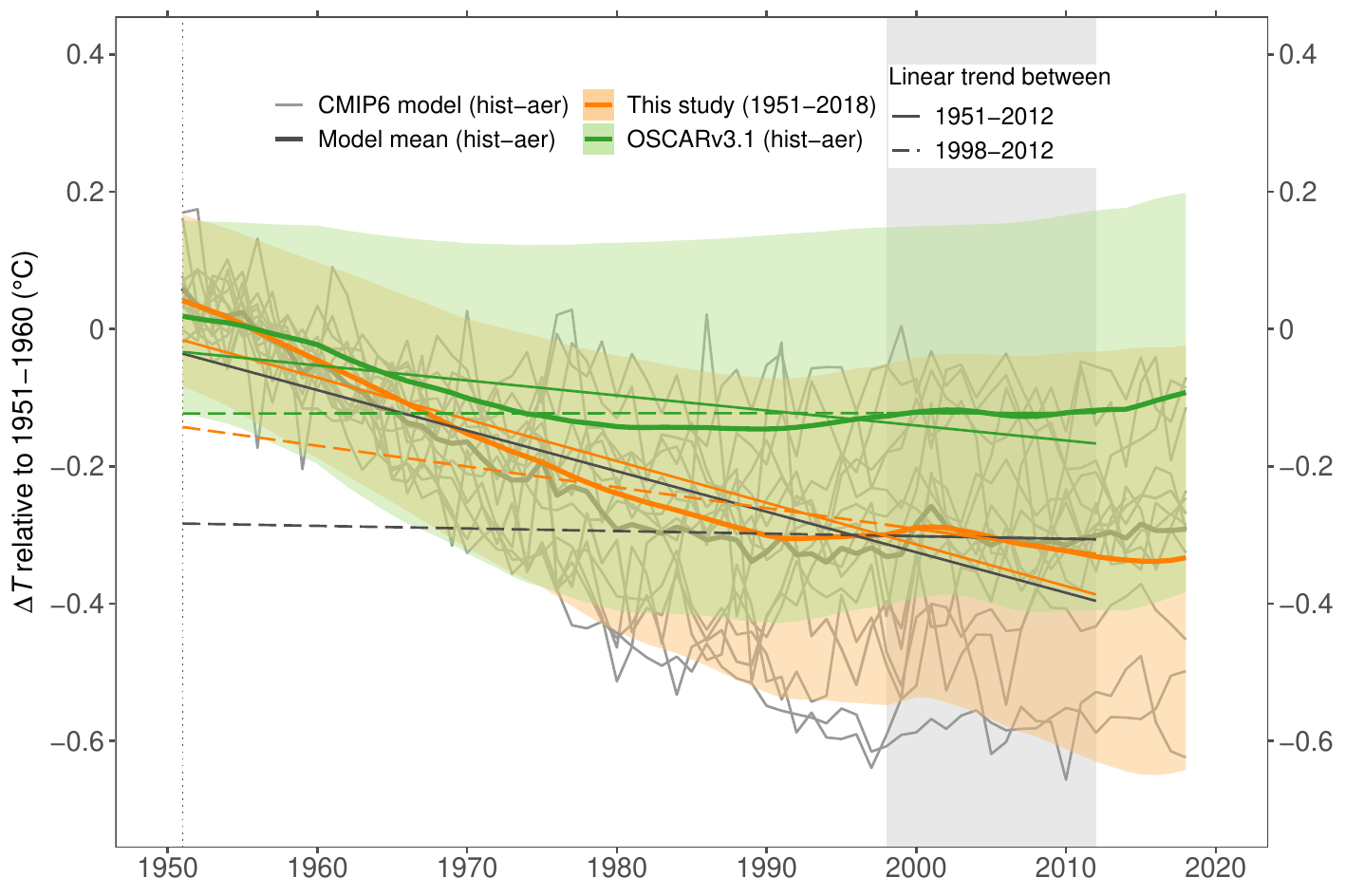}
    \caption{\textbf{Aerosol and pollutant-induced \boldmath{$\Delta T$}.} See Fig. \ref{fig:nat} for the uncertainty description. The thin solid and solid with extended dashed lines represent linear trends for the years 1951-2012 and 1998-2012, respectively. The SW era from 1998 to 2012 is shaded in light gray. See table \ref{tab:cmip6} for the investigated CMIP6 ESMs \cite{Gillett2016}.}
    \label{fig:trd_aer}
\end{figure}

\begin{figure}[ht]%
    \centering
    \includegraphics[width=0.9\textwidth]{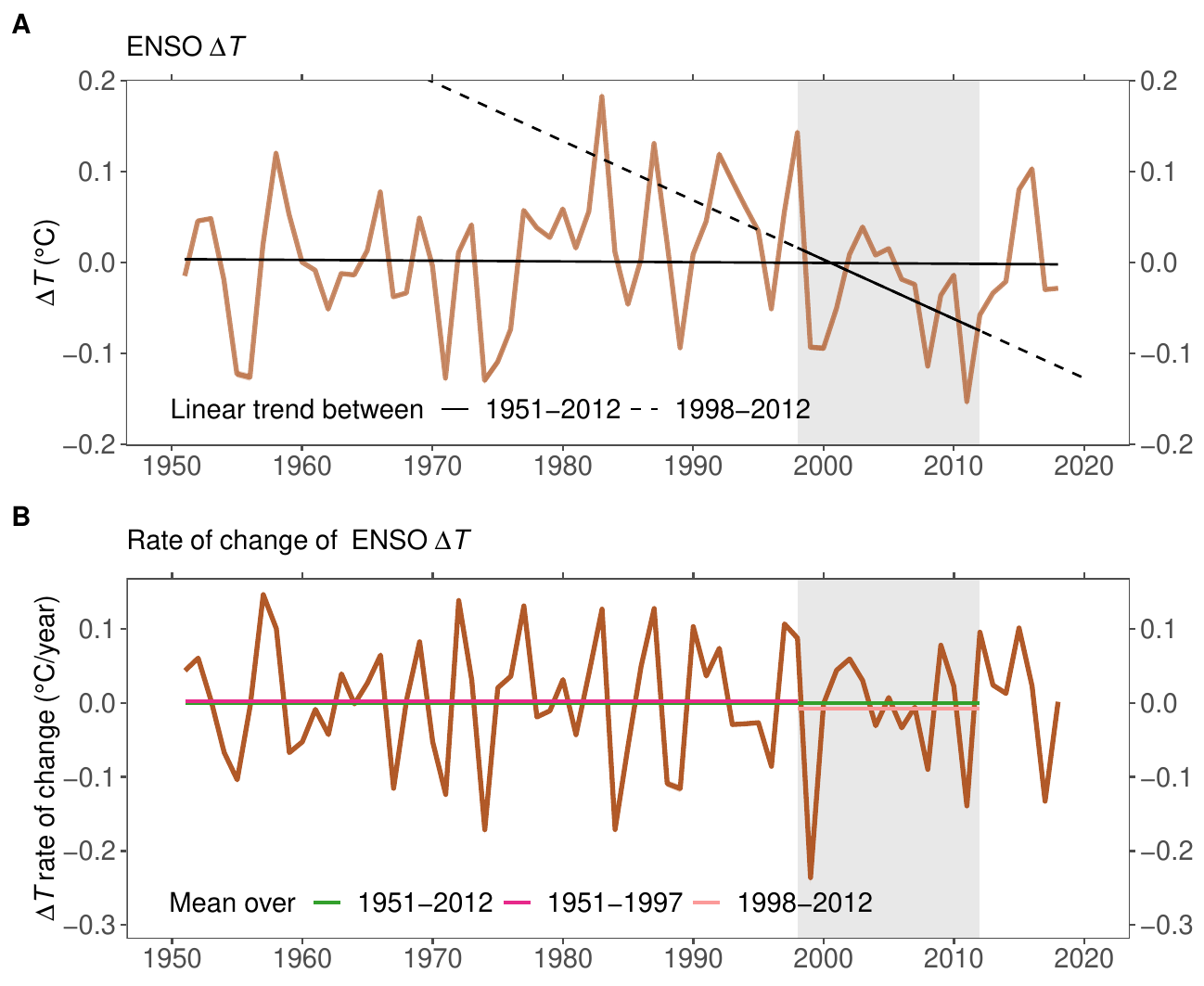}
    \caption{\textbf{\boldmath{$\Delta T$} caused by ENSO.} (\textbf{A}) \textsl{$\Delta T$} trend caused by ENSO. Linear trends across the selected periods are represented by solid and solid with extended dashed lines. (\textbf{B}) Rate of change of ENSO \textsl{$\Delta T$}. Thick horizontal lines represent the mean values for the specified periods. The MEI uses the NOAA MEI \cite{Wolter2011} (\protect\url{https://psl.noaa.gov/enso/mei.old/mei.html}), which is the default MEI in this study. The ENSO is derived from HadCRUT5. The SW era from 1998 to 2012 is shaded in light gray.}
    \label{fig:vari}
\end{figure}

\begin{figure}[ht]%
    \centering
    \includegraphics[width=0.9\textwidth]{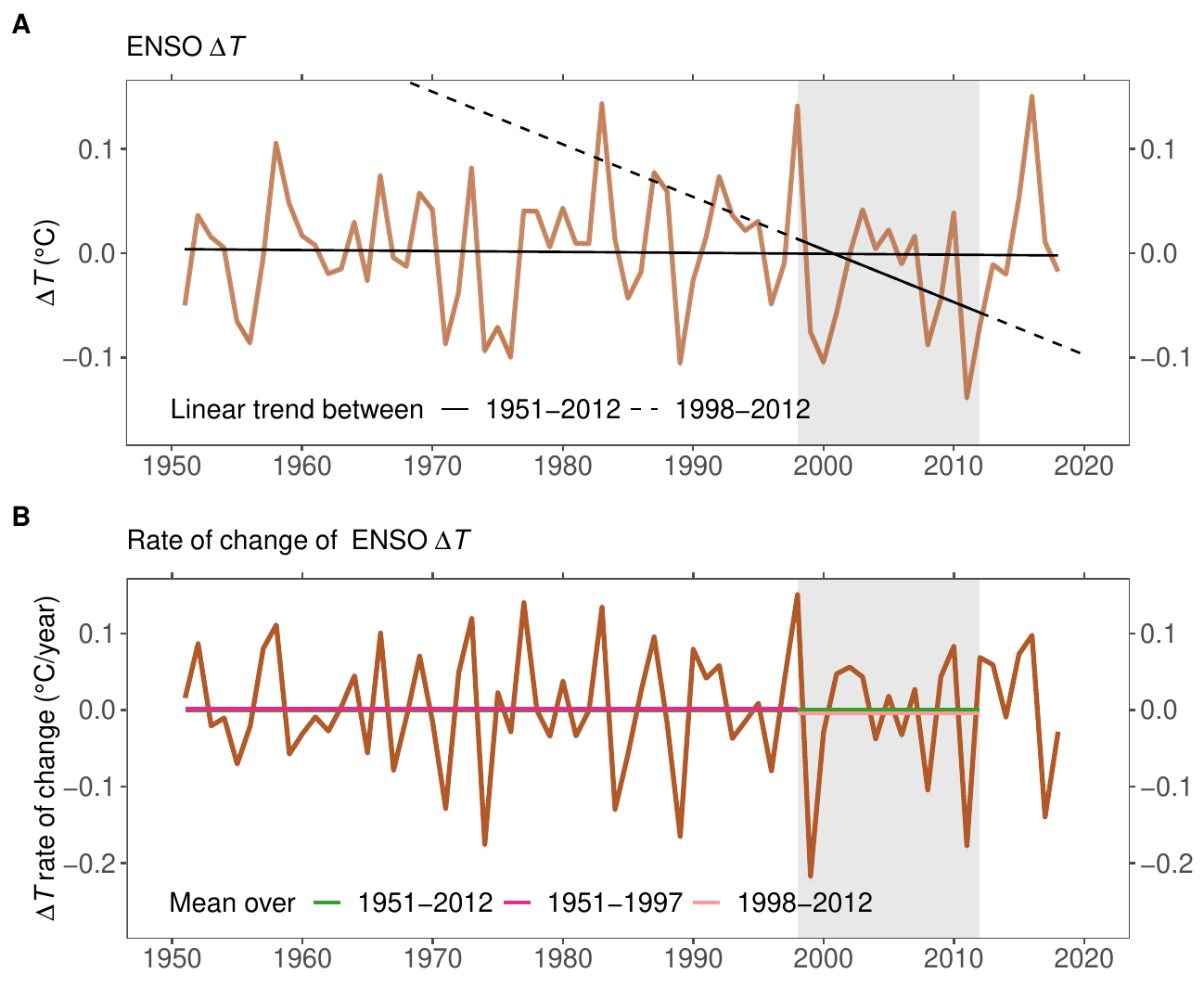}
    \caption{\textbf{\boldmath{$\Delta T$} caused by ENSO.} The same as in fig. \ref{fig:vari}, except that ENSO is estimated by using the MEI from NCEP-NCAR (\protect\url{https://www.webberweather.com/multivariate-enso-index.html}).}
    \label{fig:vari_ncep}
\end{figure}

\begin{figure}[ht]%
    \centering
    \includegraphics[width=0.9\textwidth]{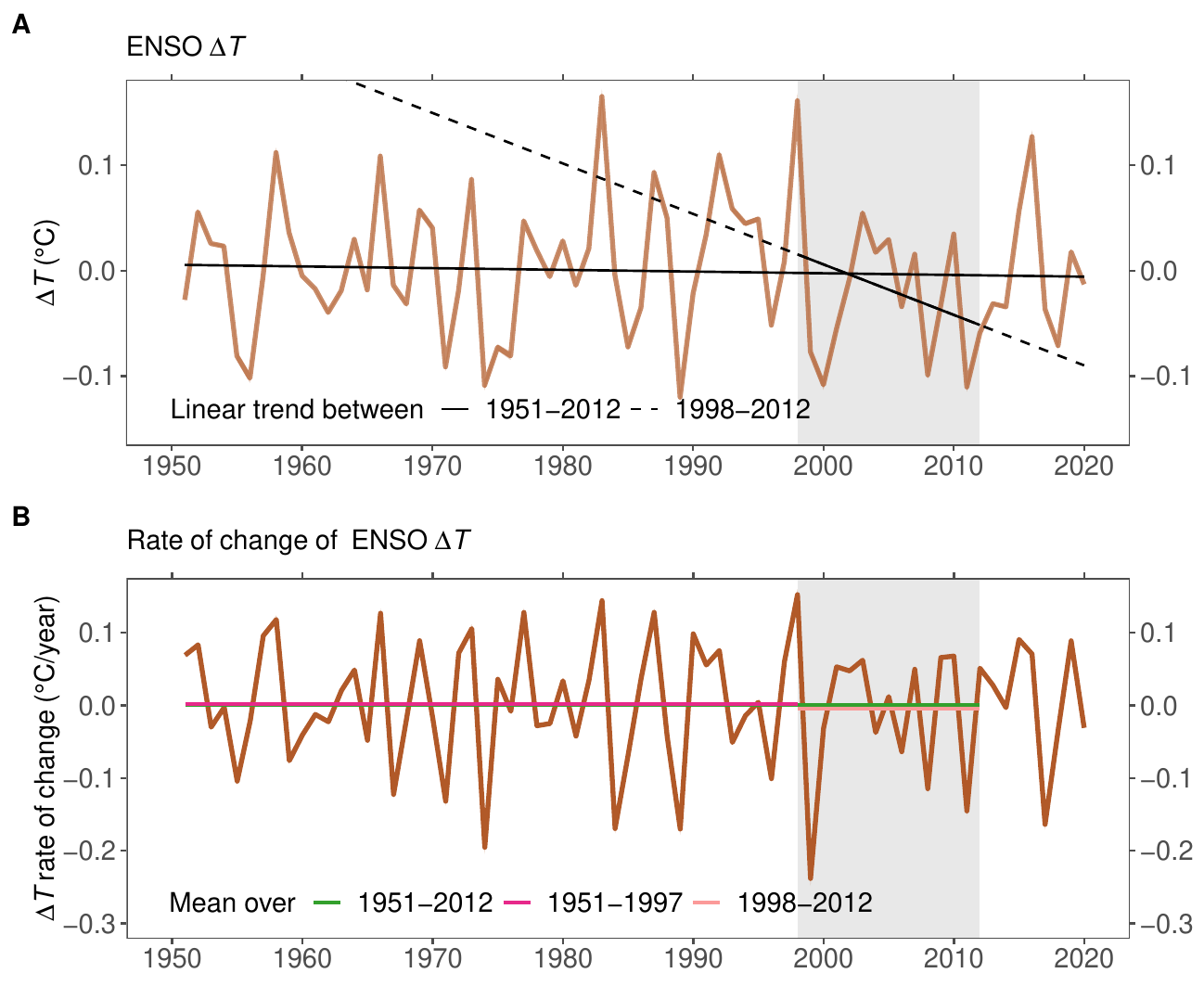}
    \caption{\textbf{\boldmath{$\Delta T$} caused by ENSO.} The same as in fig. \ref{fig:vari}, except that ENSO is estimated by using the cold tongue index (CTI) (\protect\url{https://github.com/ToddMitchellGH/Cold-Tongue-Index}).}
    \label{fig:vari_cti}
\end{figure}

\begin{figure}[ht]%
    \centering
    \includegraphics[width=0.9\textwidth]{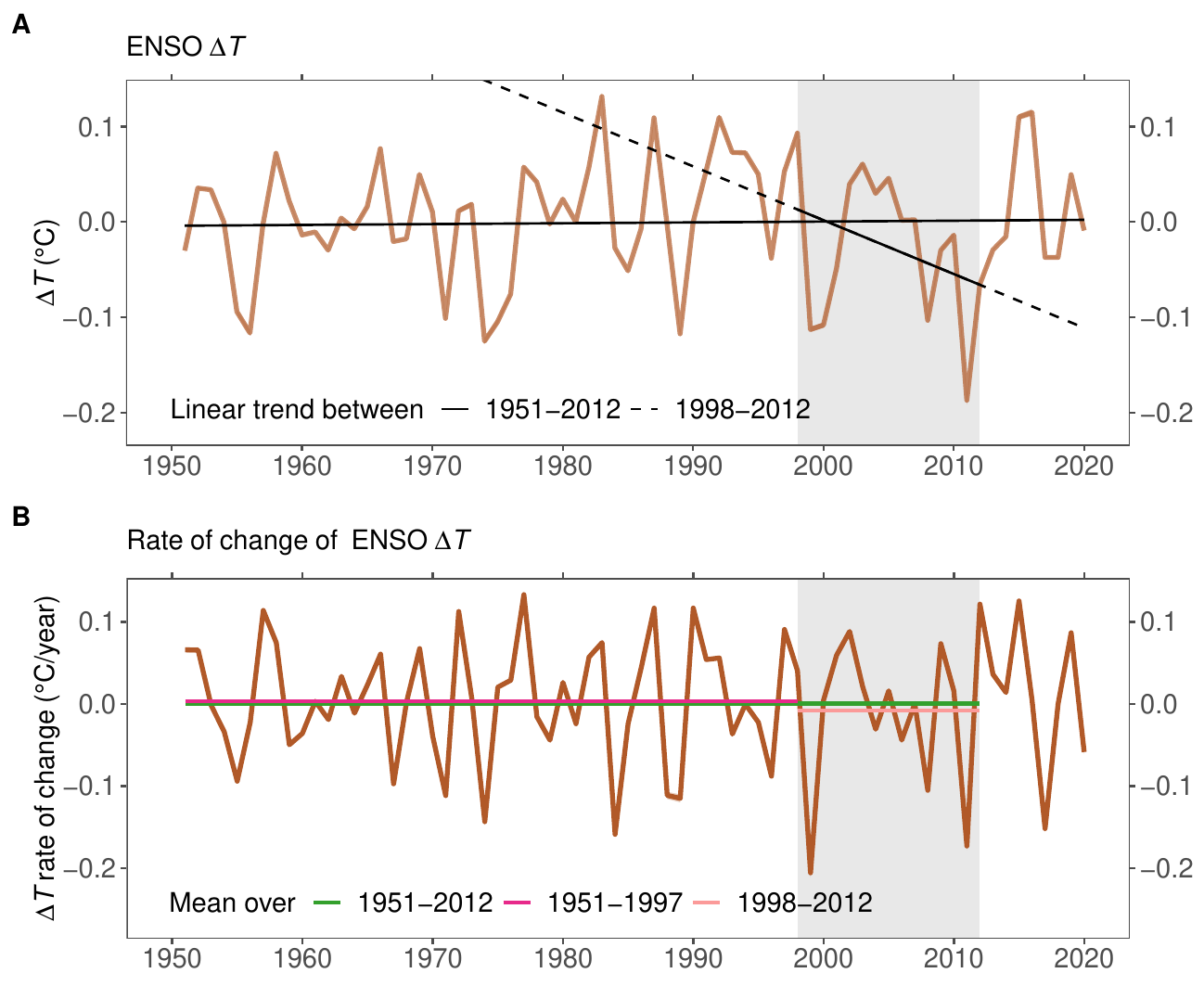}
    \caption{\textbf{\boldmath{$\Delta T$} caused by ENSO.} The same as in fig. \ref{fig:vari}, except that ENSO is estimated by using the ``BEST" ENSO Index \cite{Smith2000} (\protect\url{https://psl.noaa.gov/people/cathy.smith/best/}).}
    \label{fig:vari_best}
\end{figure}

\begin{figure}[ht]%
    \centering
    \includegraphics[width=0.9\textwidth]{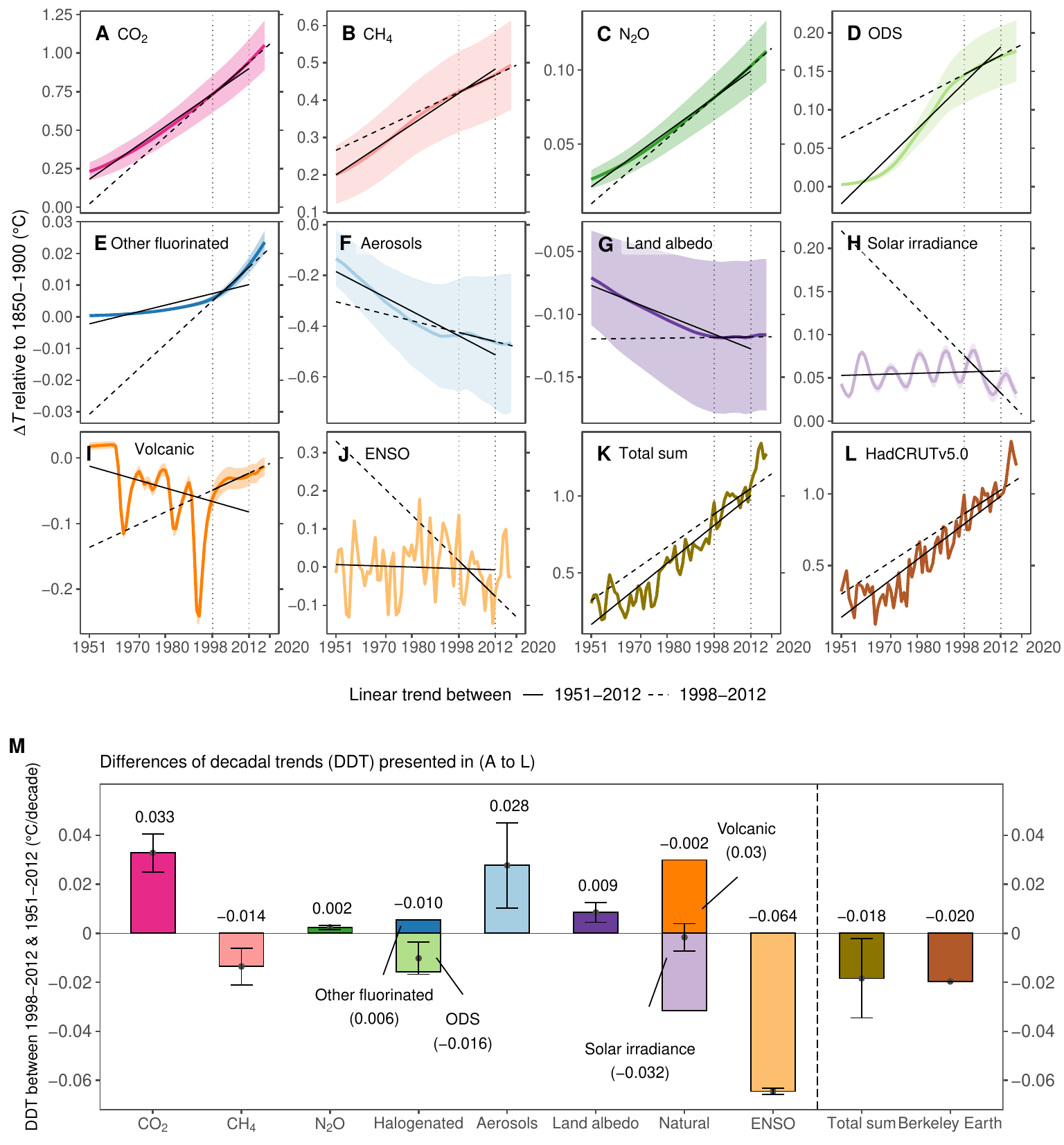}
    \caption{\textbf{Anthropogenic and natural \boldmath{$\Delta T$} by source and their DDT between 1998-2012 and 1951-2012.} The same as in Fig. \ref{fig:attrib}, except that the temperature record uses Berkeley Earth.}
    \label{fig:attrib_berk}
\end{figure}

\begin{figure}[ht]%
    \centering
    \includegraphics[width=0.9\textwidth]{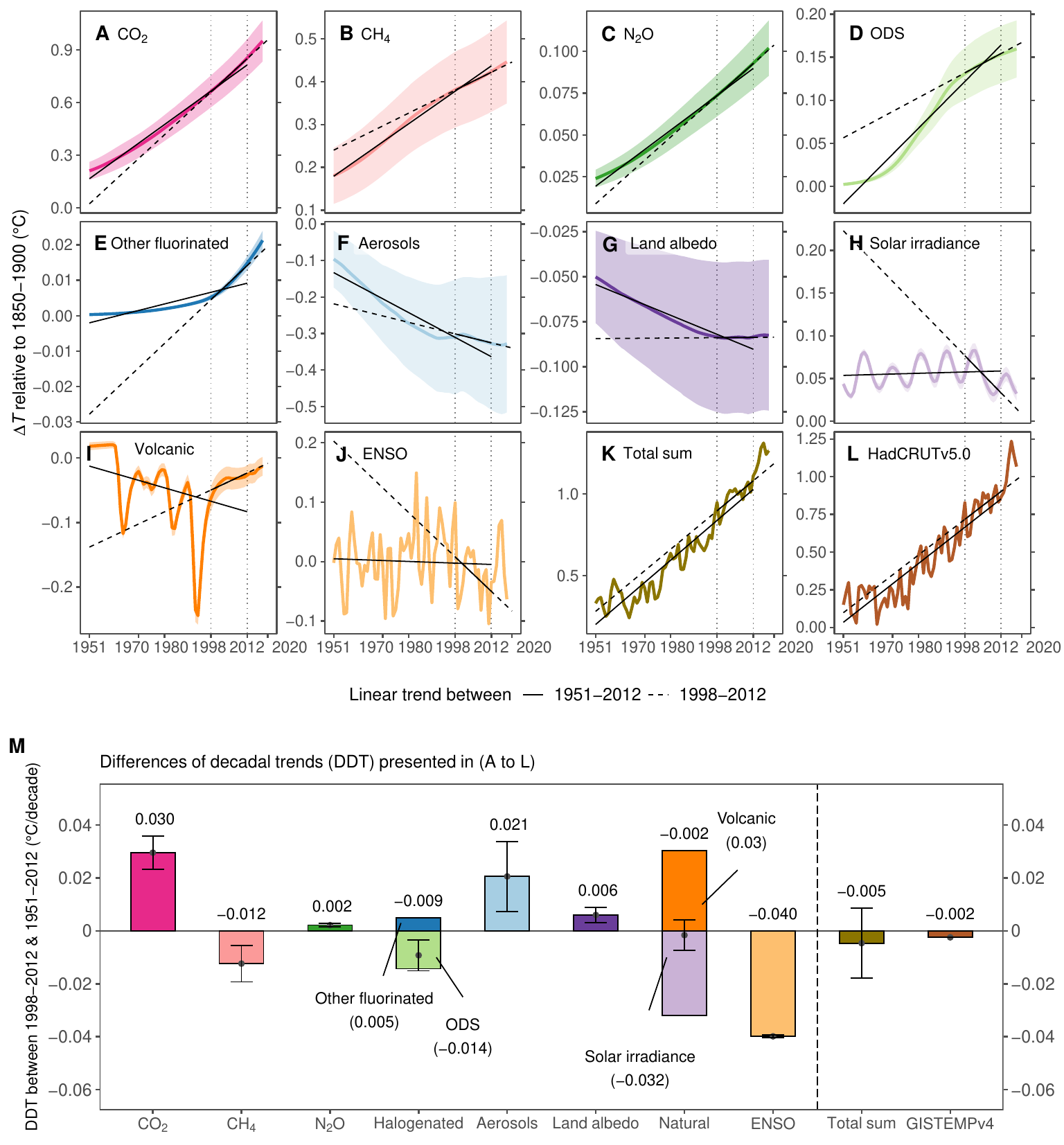}
    \caption{\textbf{Anthropogenic and natural \boldmath{$\Delta T$} by source and their DDT between 1998-2012 and 1951-2012.} The same as in Fig. \ref{fig:attrib}, except that the temperature record uses GISTEMPv4.}
    \label{fig:attrib_gis}
\end{figure}

\begin{figure}[ht]%
    \centering
    \includegraphics[width=0.9\textwidth]{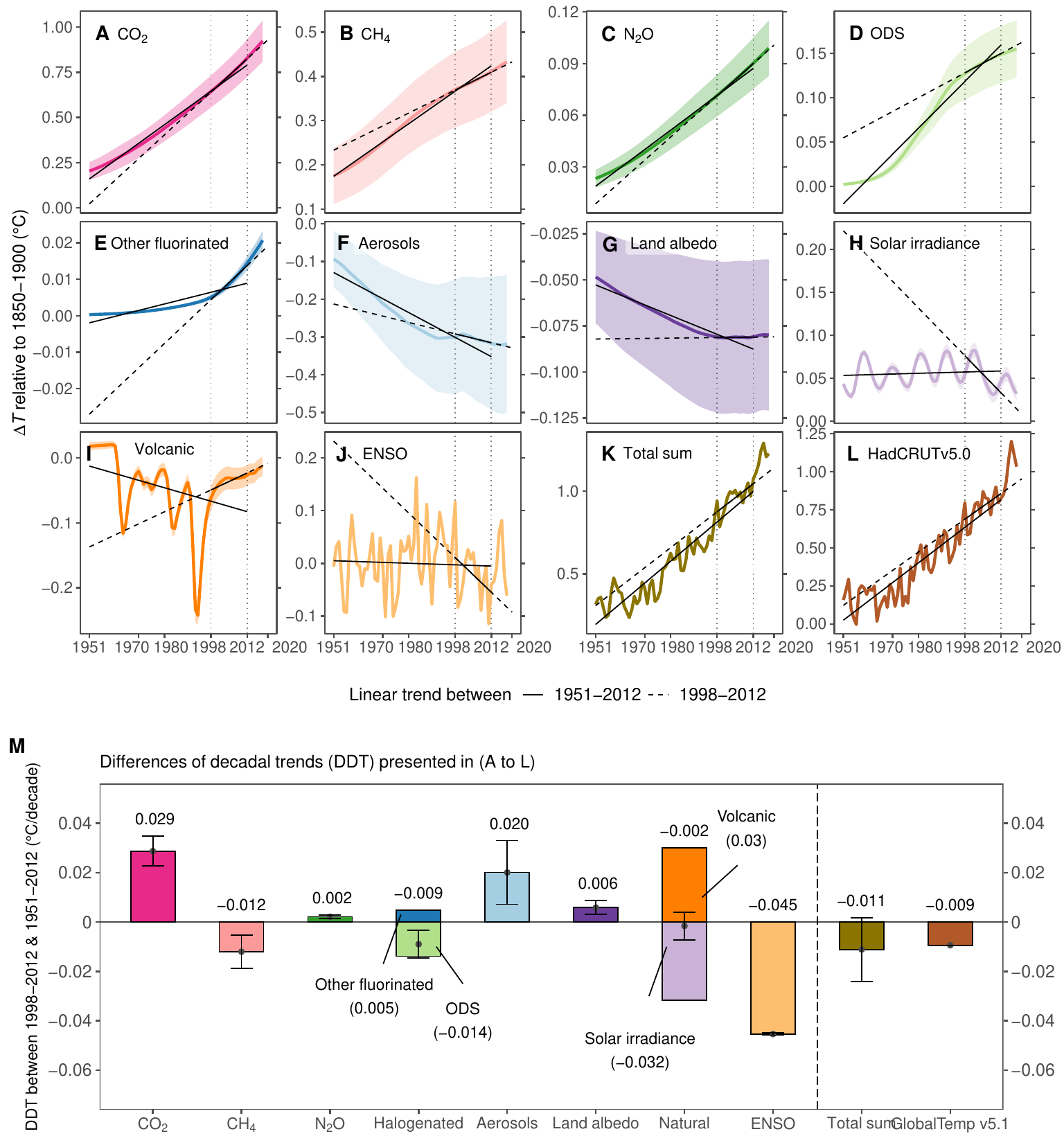}
    \caption{\textbf{Anthropogenic and natural \boldmath{$\Delta T$} by source and their DDT between 1998-2012 and 1951-2012.} The same as in Fig. \ref{fig:attrib}, except that the temperature record uses GlobalTemp v5.1.}
    \label{fig:attrib_glb}
\end{figure}

\begin{figure}[ht]%
    \centering
    \includegraphics[width=0.9\textwidth]{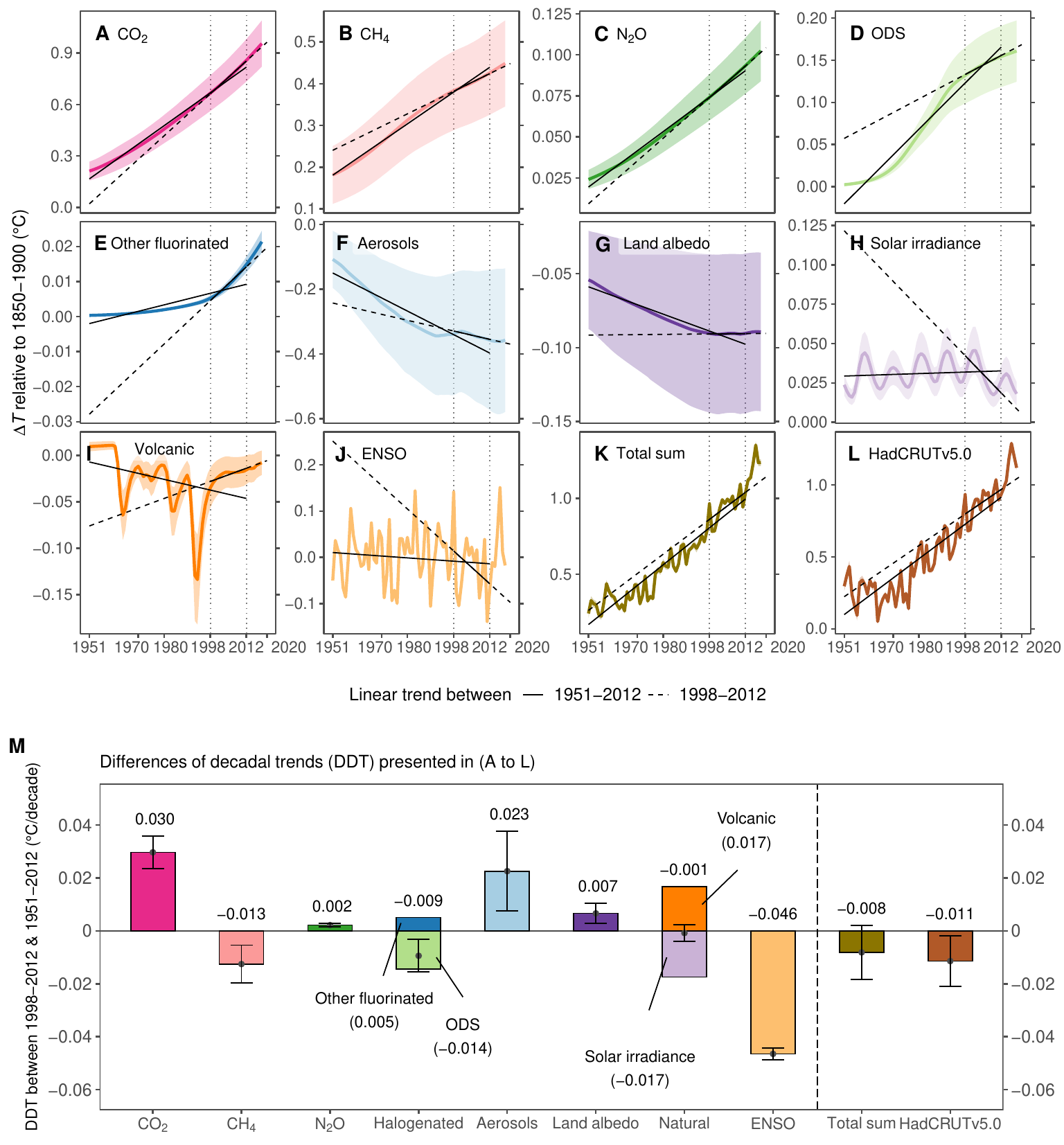}
    \caption{\textbf{Anthropogenic and natural \boldmath{$\Delta T$} by source and their DDT between 1998-2012 and 1951-2012.} The same as in Fig. \ref{fig:attrib}, except that ENSO is estimated by using the MEI from NCEP-NCAR (\protect\url{https://www.webberweather.com/multivariate-enso-index.html}).}
    \label{fig:attrib_hd50_ncep}
\end{figure}

\begin{figure}[ht]%
    \centering
    \includegraphics[width=0.9\textwidth]{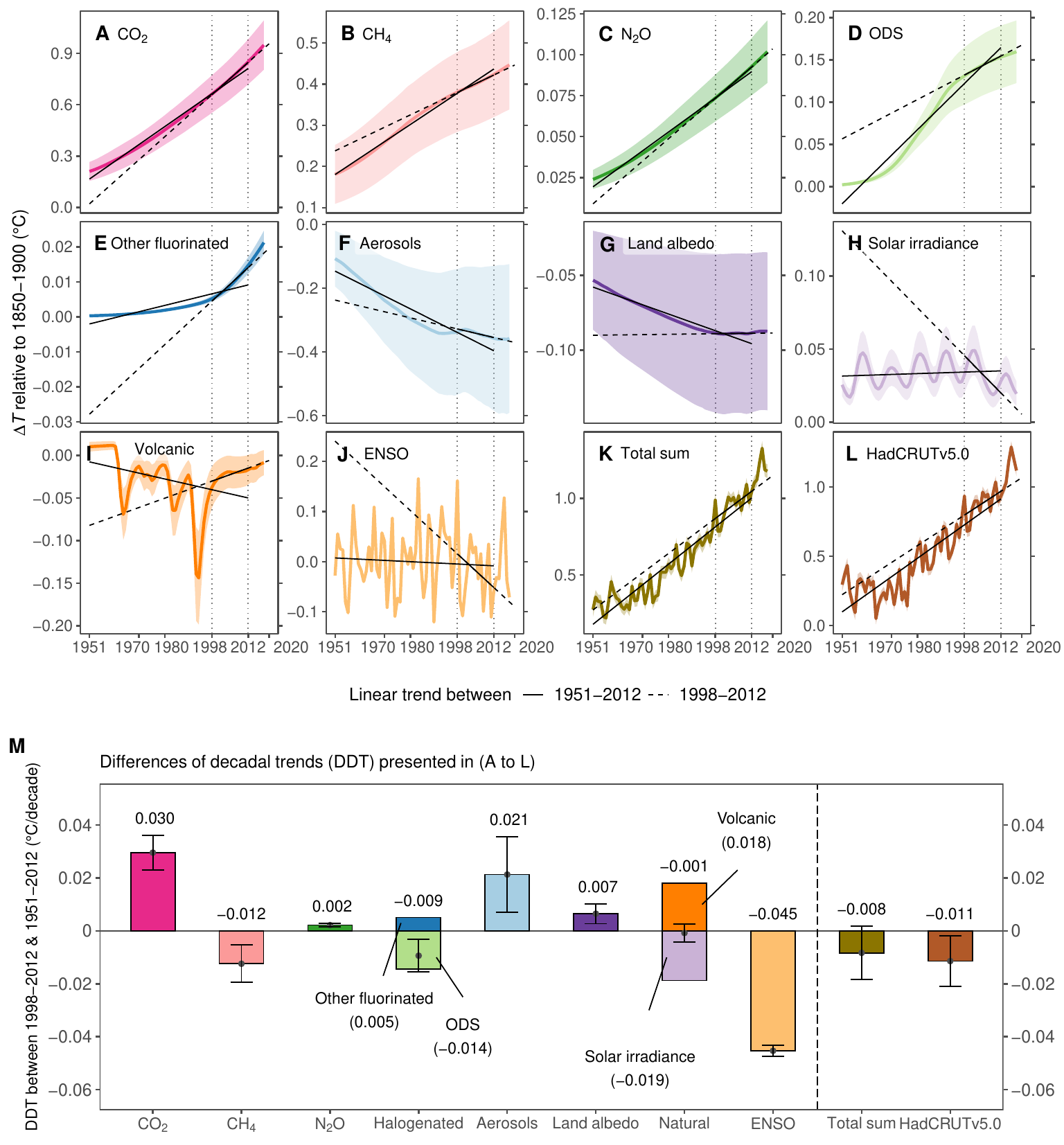}
    \caption{\textbf{Anthropogenic and natural \boldmath{$\Delta T$} by source and their DDT between 1998-2012 and 1951-2012.} The same as in Fig. \ref{fig:attrib}, except that ENSO is estimated by using the cold tongue index (CTI) (\protect\url{https://github.com/ToddMitchellGH/Cold-Tongue-Index}).}
    \label{fig:attrib_hd50_cti}
\end{figure}

\begin{figure}[ht]%
    \centering
    \includegraphics[width=0.9\textwidth]{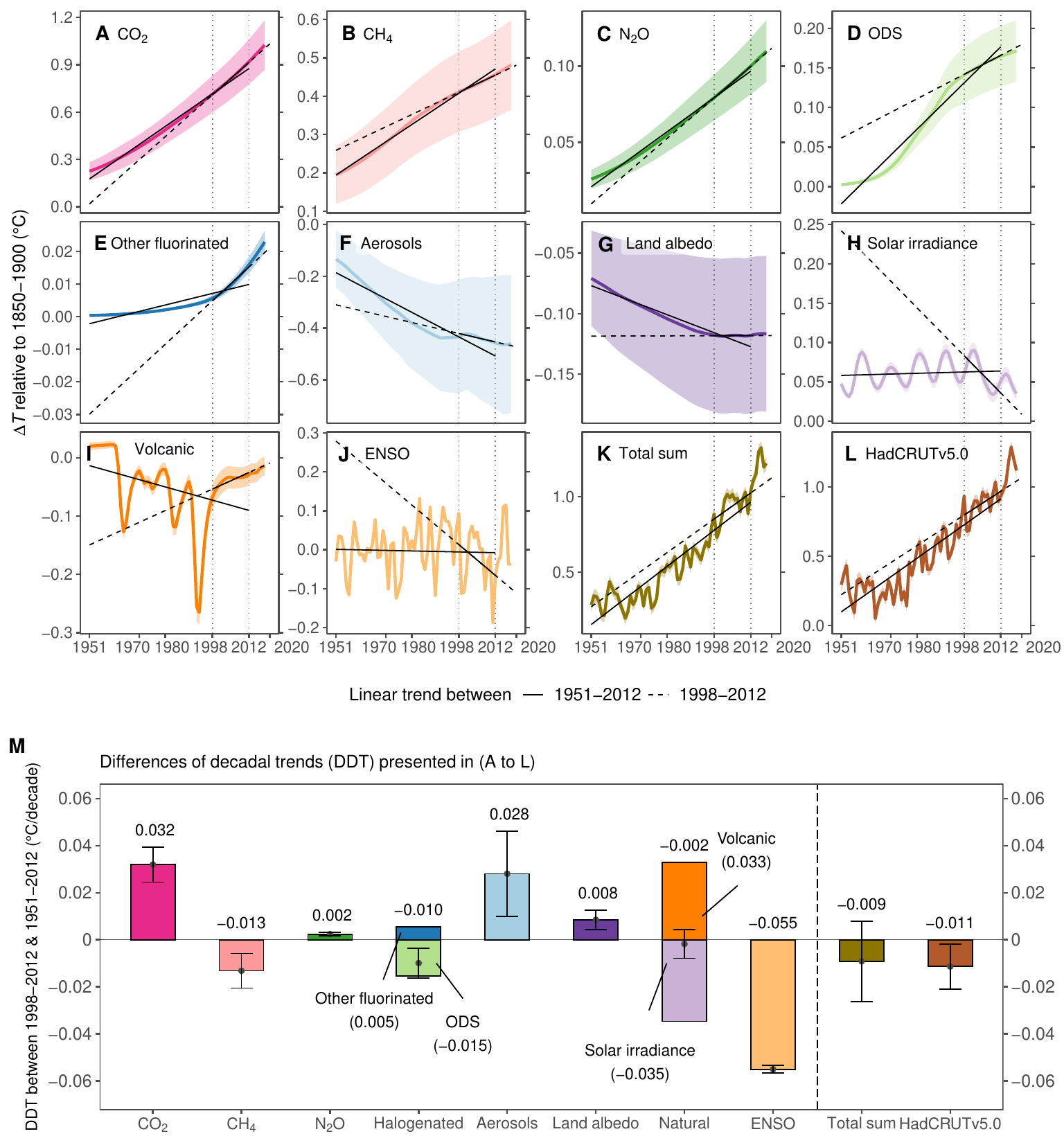}
    \caption{\textbf{Anthropogenic and natural \boldmath{$\Delta T$} by source and their DDT between 1998-2012 and 1951-2012.} The same as in Fig. \ref{fig:attrib}, except that ENSO is estimated by using the ``BEST" ENSO Index \cite{Smith2000} (\protect\url{https://psl.noaa.gov/people/cathy.smith/best/}).}
    \label{fig:attrib_hd50_best}
\end{figure}

\begin{figure}[ht]%
    \centering
    \includegraphics[width=0.9\textwidth]{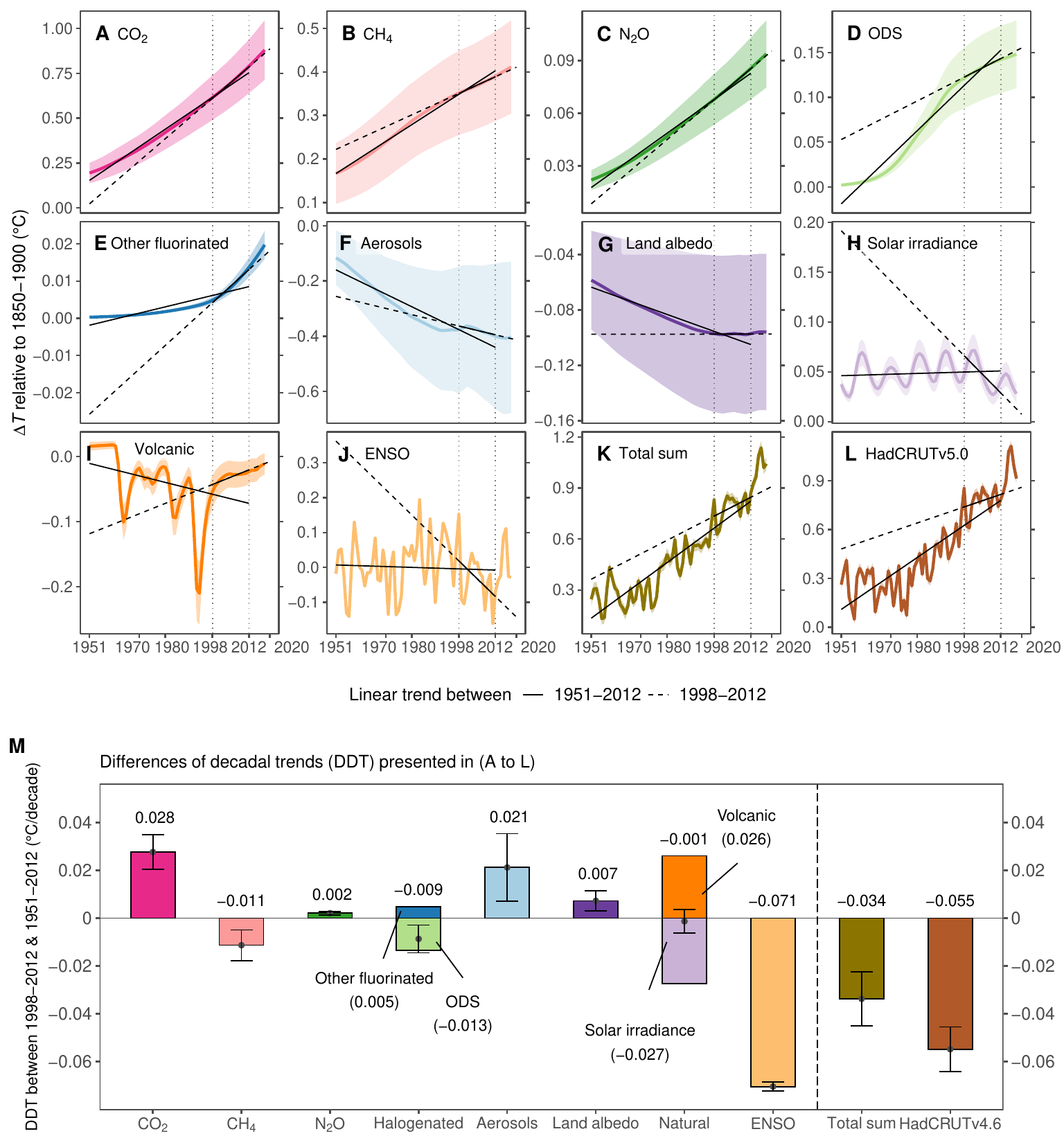}
    \caption{\textbf{Anthropogenic and natural \boldmath{$\Delta T$} by source and their DDT between 1998-2012 and 1951-2012.} The same as in Fig. \ref{fig:attrib}, except that the temperature record uses HadCRUT4.6.}
    \label{fig:attrib_hd46}
\end{figure}

\noindent Materials and Methods \\
Figs. \ref{fig:tem_obsrv} to \ref{fig:attrib_my_hd50}\\
Tables \ref{tab:bic} to \ref{tab:coef}\\
References \textit{(56-64)}

\pagebreak    
\clearpage

\section*{Materials and Methods}
\subsection*{\boldmath{$\Delta T$} simulations}

\begin{figure}[ht]%
    \centering
    \includegraphics[width=0.9\textwidth]{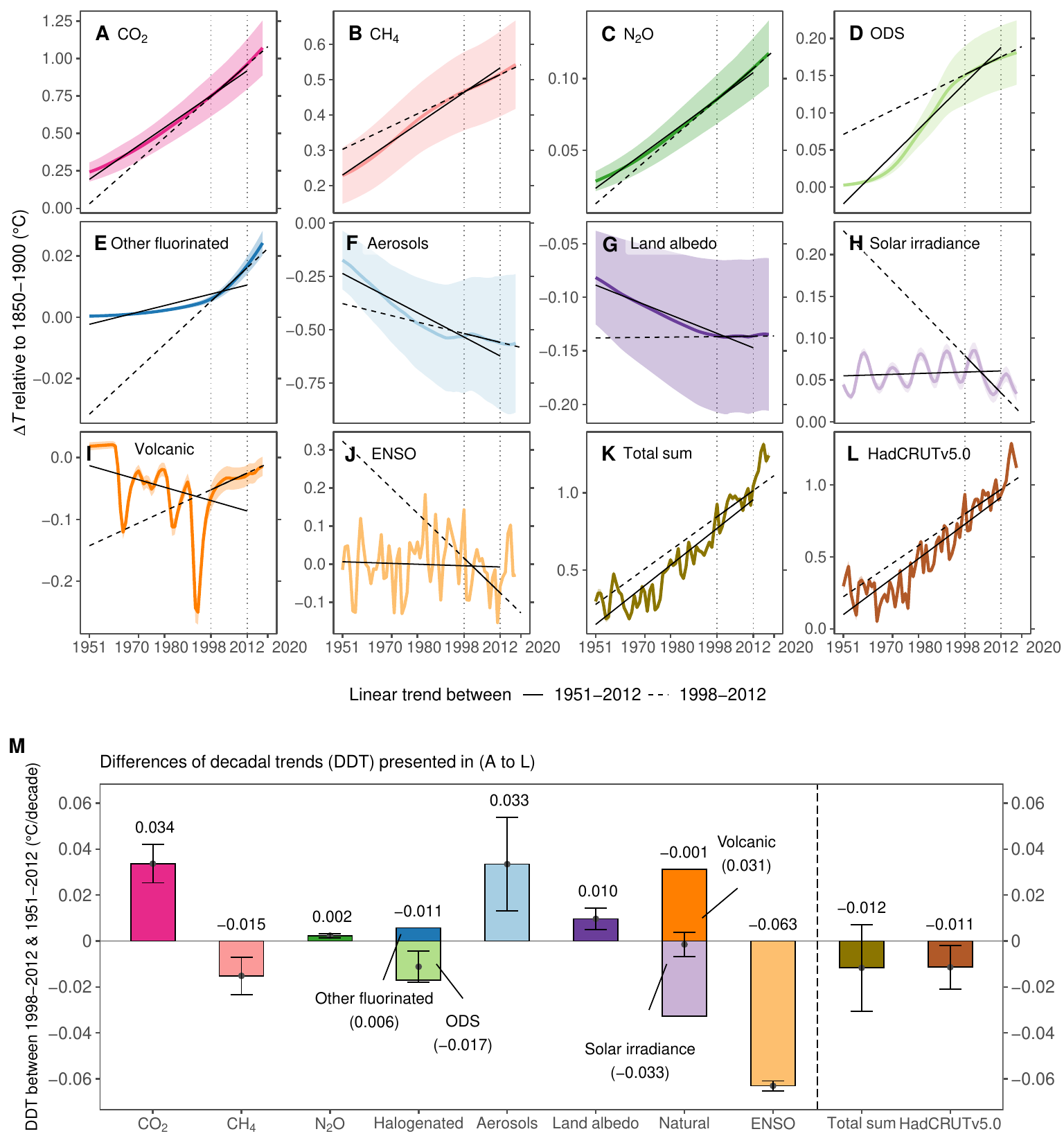}
    \caption{\textbf{Anthropogenic and natural \boldmath{$\Delta T$} by source and their DDT between 1998-2012 and 1951-2012.} The same as in Fig. \ref{fig:attrib}, except that the methane forcing is parameterized using the Etminan method \cite{Etminan2016}.}
    \label{fig:attrib_et_hd50}
\end{figure}

\begin{figure}[ht]%
    \centering
    \includegraphics[width=0.9\textwidth]{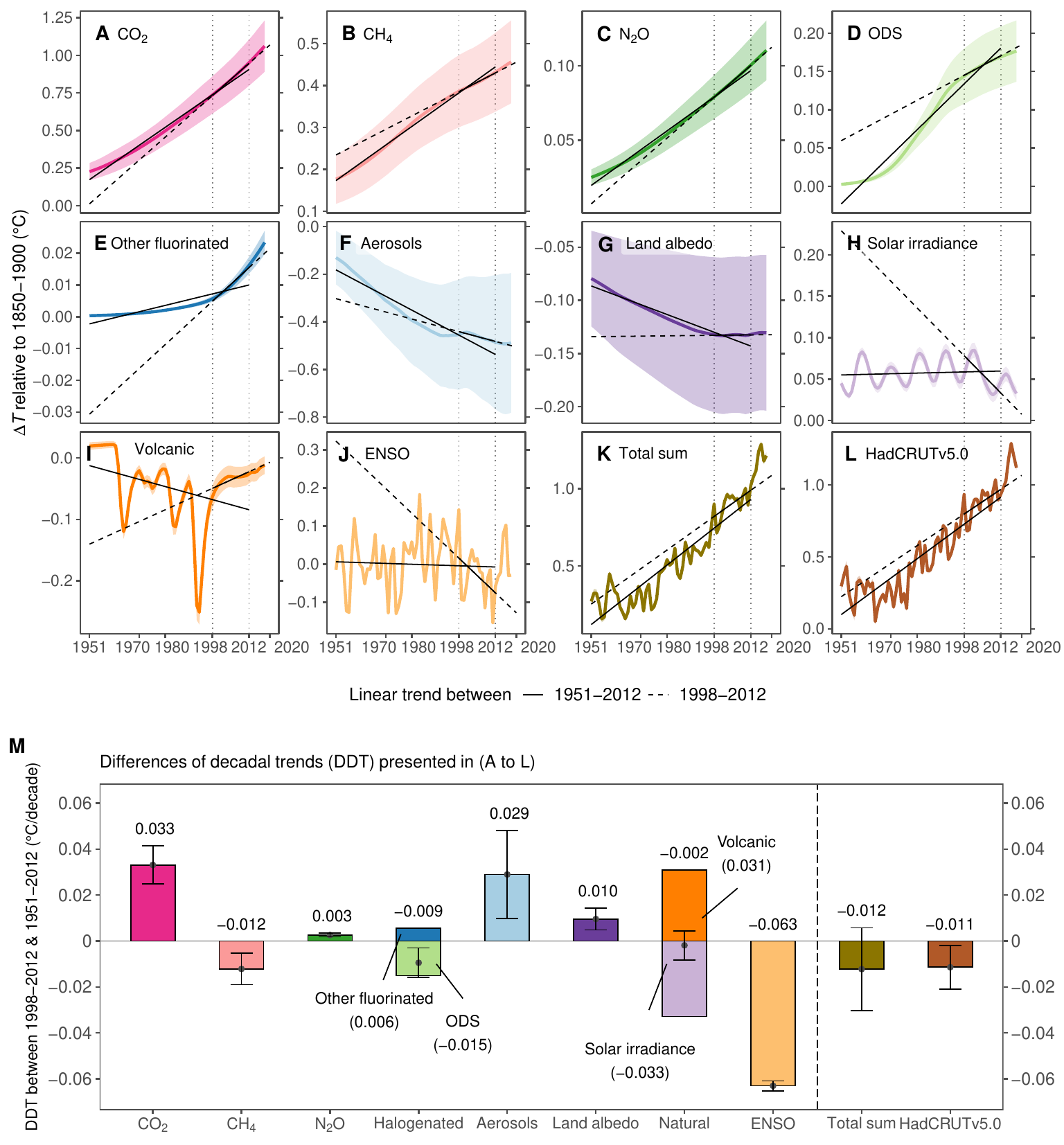}
    \caption{\textbf{Anthropogenic and natural \boldmath{$\Delta T$} by source and their DDT between 1998-2012 and 1951-2012.} The same as in Fig. \ref{fig:attrib}, except that the methane forcing is parameterized using the usual IPCC method \cite{ar5ch8}.}
    \label{fig:attrib_my_hd50}
\end{figure}

We used an RCM - SCM4OPT v3.3 \cite{Su2017, Su2018, Nicholls2021, Su2022} to simulate global \textsl{$\Delta T$} trends resulting from anthropogenic factors (such as CO\textsubscript{2}, CH\textsubscript{4}, N\textsubscript{2}O, halogenated gases (16 ODS and 23 other fluorinated gases), carbon monoxide (CO), nitrogen oxide (NO\textsubscript{x}), volatile organic compounds (VOCs), sulfate (SO\textsubscript{x}), black carbon (BC), organic carbon (OC), and land albedo) and natural forcings (such as volcanic and solar irradiance). SCM4OPT v3.3 was created by updating an older version, SCM4OPT v3.2 \cite{Su2022}. We present a new parameterization for CH\textsubscript{4} forcing \cite{Etminan2016}. The methane \textsl{$\Delta T$} trend in this study was estimated using a hybrid of the new Etminan parameterization and a traditional technique as in AR5 \cite{ar5ch8}, and the associated processes and parameters were used in the Monte Carlo simulation as below. The former showed a 25\% higher CH\textsubscript{4} forcing than the latter (1750–2011), resulting in a lower DDT level of -0.015$\pm$0.008~$^{\circ}$C/decade (cf. -0.012$\pm$0.007~$^{\circ}$C/decade using AR5's method) in the final attribution (figs. \ref{fig:attrib_et_hd50}-\ref{fig:attrib_my_hd50}). 

A normalized marginal method \cite{Trudinger2005, Li2016, Fu2021, Boucher2021} was applied to SCM4OPT v3.3 for \textsl{$\Delta T$} quantifications of climate forcers. \textsl{$\Delta T$} trends induced by nonemission factors such as natural forcings including volcanoes and solar irradiance, as well as human factors such as land albedo resulting from land-use change, were quantified using a residual method: (1) one historical emulation with all emissions and nonemission factors as input; (2) one ``exclusive'' experiment for estimating land-use albedo \textsl{$\Delta T$} trends with constant preindustrial land cover levels, and two ``exclusive'' experiments for estimating natural forcing \textsl{$\Delta T$} trends by deleting volcanic or solar forcings from historical emulations. We compared \textsl{$\Delta T$} obtained from historical and ``exclusive'' experiments, and the differences between the two experiments were \textsl{$\Delta T$} trends caused by these nonemission factors. For the remaining \textsl{$\Delta T$}, the normalized marginal method was used to quantify the emission-induced \textsl{$\Delta T$} trends \cite{Su2022}. We assumed that the ratio of a species' ($e$) \textsl{$\Delta T$} to overall \textsl{$\Delta T$}, defined as $\alpha^{t}_{e}$, is proportional to the marginal effect of $e$ divided by the total marginal effect. To estimate $\alpha^{t}_{e}$, two simulations were carried out for each species: (1) one historical simulation as above and the resulting \textsl{$\Delta T$} denoted as $\Delta T_{all}$; and (2) another identical simulation except emission $e$ reduced by a fraction of $\epsilon$ = 0.001 to calculate the \textsl{$\Delta T$} termed as $\Delta T_{e,\epsilon}$. Therefore, the relative contribution $\alpha^{t}_{e}$ can be obtained:

\begin{equation} \label{relcon}
    \alpha^{t}_{e} = \frac{\Delta T_{all}-\Delta T_{e,\epsilon}}{\sum_{e\prime}{\left(\Delta T_{all}-\Delta T_{e\prime,\epsilon}\right)}}
\end{equation}

$\Delta T_{e}$ resulting from emission $e$ can be calculated as follows:

\begin{equation}
\label{eq:marginal2}
\Delta T_{e} = {\Delta T_{all}} \cdot \alpha^{t}_{e}
\end{equation}

After calculating \textsl{$\Delta T$} induced by emissions and nonemission factors, we classified the resulting \textsl{$\Delta T$} as CO\textsubscript{2}, CH\textsubscript{4}, N\textsubscript{2}O, ODS, other fluorinated gases, aerosols and pollutants, land albedo, volcanoes, and solar irradiance, considering their inherent characteristics and rates of change since the preindustrial revolution.

\begin{figure}[ht]%
    \centering
    \includegraphics[width=0.9\textwidth]{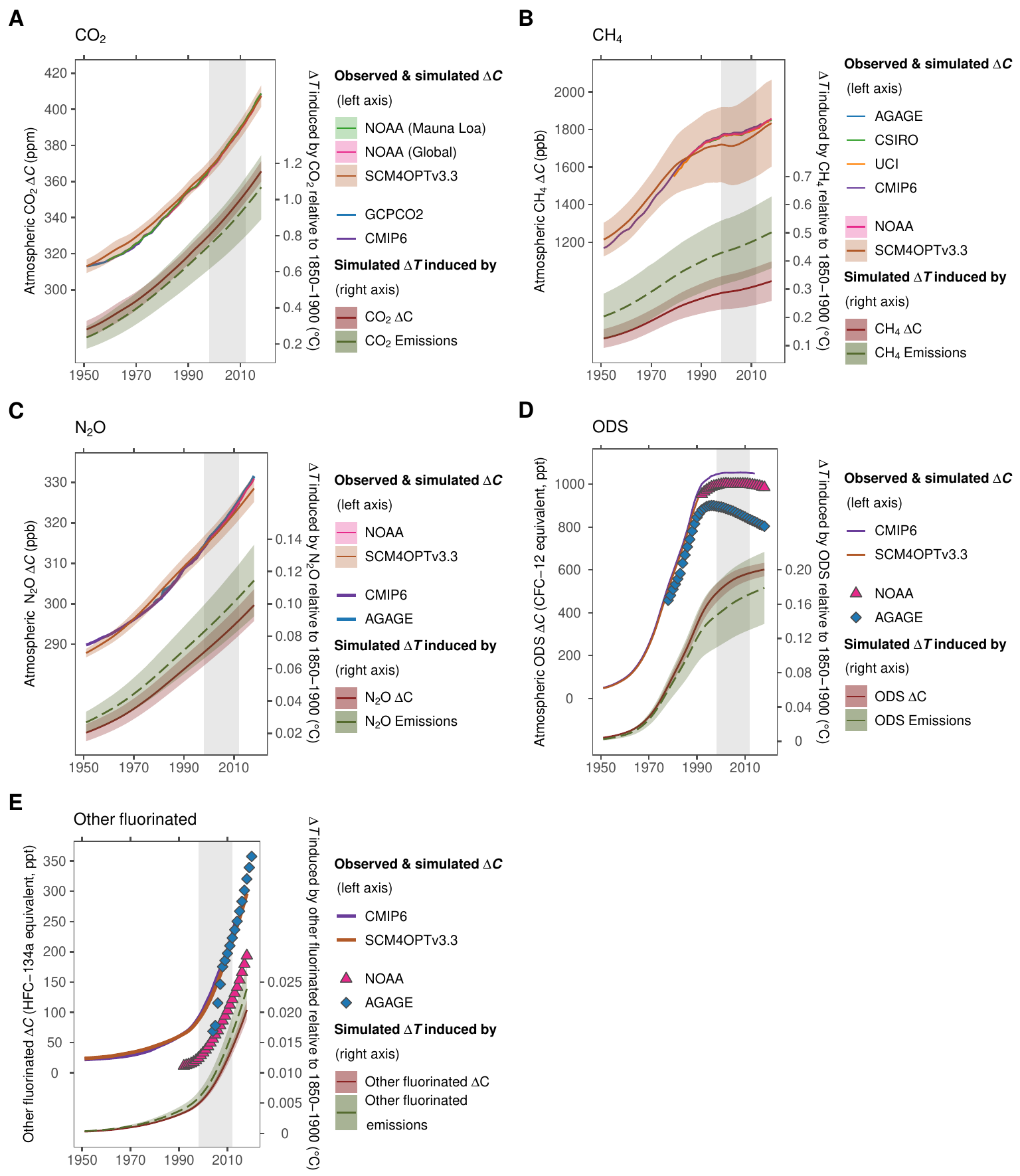}
    \caption{\textbf{Observed and simulated atmospheric \boldmath{$\Delta C$} and the induced \boldmath{$\Delta T$}.} See Fig. \ref{fig:cause} for the uncertainty description. The components of ODS and other fluorinated gases for each source in (D and E) differ (table \ref{tab:comp}). Uncertainties are not considered in the \textsl{$\Delta C$} of ODS and other fluorinated gases because they are equivalent values combining a collection of compounds \cite{Meinshausen2017}. The SW era from 1998 to 2012 is shaded in light gray.}
    \label{fig:conc_trd}
\end{figure}

To estimate \textsl{$\Delta T$} trends induced by GHGs \textsl{$\Delta C$} (Fig. \ref{fig:cause}, fig. \ref{fig:conc_trd}), we again used the above normalized marginal method for the forcing-driven mode of SCM4OPT v3.3. We calculated the associated \textsl{$\Delta F$} from \textsl{$\Delta C$} and applied $\epsilon$ = 0.001 to \textsl{$\Delta F$} to obtain the trend of \textsl{$\Delta C$}-induced \textsl{$\Delta T$}. The same $\beta_{GHG}$ (as below) was multiplied to estimate the final \textsl{$\Delta T$} induced by the GHG \textsl{$\Delta C$}.

It is worth noting that the numbers of ODS or other fluorinated gases used in this study differ from the compiled data or observations (table \ref{tab:comp}). For example, a few ODS gases (such as CFC-13, CH\textsubscript{2}Cl\textsubscript{2} and CHCl\textsubscript{3}) and other fluorinated gases (such as SO\textsubscript{2}F\textsubscript{2}, C\textsubscript{7}F\textsubscript{16}, C\textsubscript{8}F\textsubscript{18} and C\textsubscript{2}Cl\textsubscript{4}) were not included in our simulation due to data availability. However, the warming effects from these gases are relatively small, and our results retained almost the same total equivalent \textsl{$\Delta C$} trends as the observations (Fig. \ref{fig:cause}, fig. \ref{fig:conc_trd}). Therefore, our conclusions are unaffected.

\subsection*{Statistical model}

ENSO has larger influences on the global \textsl{$\Delta T$} than other internal variabilities \cite{Lean2008,Thompson2009,Foster2011,Lean2018}. We used ENSO to reflect the main changes due to internal variability. SCM4OPT v3.3 does not replicate ENSO. We employed a statistical model to correct the biases in the magnitude of the simulated \textsl{$\Delta T$}, and distinguish ENSO influences from the observed \textsl{$\Delta T$} \cite{Lean2008,Foster2011,Lean2018}. We hypothesized that the observed \textsl{$\Delta T$} consists of \textsl{$\Delta T$} induced by anthropogenic GHG emissions, other anthropogenic factors such as aerosols and pollutants and land albedo, natural forcings, and ENSO. Accordingly, the observed \textsl{$\Delta T$} is defined as a multiple regression among the time series of these elements plus a residual component allowing for errors resulting from simulations and observations:

\begin{equation} \label{statmdl}
    \begin{split}
    \Delta T_{obs}(t)  & = \beta_{GHG}(\Delta T_{CO_{2}}(t) + \Delta T_{CH_{4}}(t) + \Delta T_{N_{2}O}(t) + \Delta T_{ODS}(t) + \Delta T_{oFGS}(t)) \\
                       & + \beta_{non-GHG}(\Delta T_{aero}(t) + \Delta T_{lcc}(t)) \\
                       & + \beta_{nat}(\Delta T_{volc}(t) + \Delta T_{solar}(t)) \\
                       & + \beta^{1}_{MEI}MEI(t-\tau_{1}) + \beta^{2}_{MEI}MEI(t-\tau_{2}) + \beta^{3}_{MEI}MEI(t-\tau_{3}) \\
                       & + Resi(t) \\
                       & + T_{0}
    \end{split}
\end{equation}

where $\Delta T_{obs}(t)$ is the observed \textsl{$\Delta T$}. $\Delta T_{CO_{2}}(t)$, $\Delta T_{CH_{4}}(t)$, $\Delta T_{N_{2}O}(t)$, $\Delta T_{ODS}(t)$, $\Delta T_{oFGS}(t)$, $\Delta T_{aero}(t)$, $\Delta T_{lcc}(t)$, $\Delta T_{volc}(t)$ and $\Delta T_{solar}(t)$ indicate the simulated \textsl{$\Delta T$} caused by CO\textsubscript{2}, CH\textsubscript{4}, N\textsubscript{2}O, ODS, other fluorinated gases, aerosols and pollutants, land albedo, volcanoes and solar irradiance, respectively. $\beta_{GHG}$, $\beta_{non-GHG}$ and $\beta_{nat}$ are fitted coefficients for the associated factors. $MEI(t)$ indicates the monthly multivariate ENSO index (MEI) taken from existing literature \cite{Wolter2011}. Before applying the MEI, we eliminated the linear trend from it. $\beta^{i}_{MEI}$ and $\tau_{i}$ ($i=1,2,3$) represent the fitted coefficients and the delayed months for the MEI, respectively, which were selected to maximize the variance explained (table \ref{tab:coef}). $Resi(t)$ and $T_{0}$ are the residuals and intercept, respectively. The time series of the \textsl{$\Delta T$} trends estimated from SCM4OPT v3.3 are bimonthly data, which we linearly interpolated into monthly values before employing in the statistical model. Multiple regression was applied to data from 1951 to 2018. The fitted coefficients, $R^{2}$ values and correlation coefficients are shown in table \ref{tab:coef}. 

\subsection*{Uncertainties}

This study considered uncertainties in historical emissions, climate model uncertainties, and uncertainties found in various observation records. First, three emission datasets, i.e., the Community Emissions Data System (CEDS) \cite{Hoesly2018}, Emission Database for Global Atmospheric Research (EDGAR) v7.0\_GHG 1970-2021 and EDGAR v6.1\_AP 1970-2018 \cite{edgar_ghg2021,edgar_co22022}, and ACCMIP \cite{Lamarque2013}, were used to represent uncertainties resulting from estimates of historical emissions. We compiled and processed the data in the same way as ref. \cite{Su2022}. When a certain species was not available, we used equivalent emission data from a separate dataset. These emissions were used to generate the atmospheric \textsl{$\Delta C$} and the associated \textsl{$\Delta F$}, and then to estimate the global \textsl{$\Delta T$}. Second, we performed a Monte Carlo simulation with $n = 1000$ (see table S4 in the forcing study \cite{Su2022} for the parameter sets) for the climate system to calculate the climate model uncertainties. Third, we carried out regressions for individual temperature records or the representative record HadCRUT5 with different ENSO indices, considering that ENSO effects and residuals are somewhat sensitive to the temperature records and ENSO indices. The statistical regression was also performed using the Monte Carlo method with $n = 1000$. For example, the observed value was randomly chosen from the available dataset ensemble (for HadCRUT5 $n = 200$), while the predictor variables were randomly selected from the ensemble of SCM4OPTv3.3 simulations.   

\begin{figure}[ht]%
    \centering
    \includegraphics[width=0.75\textwidth]{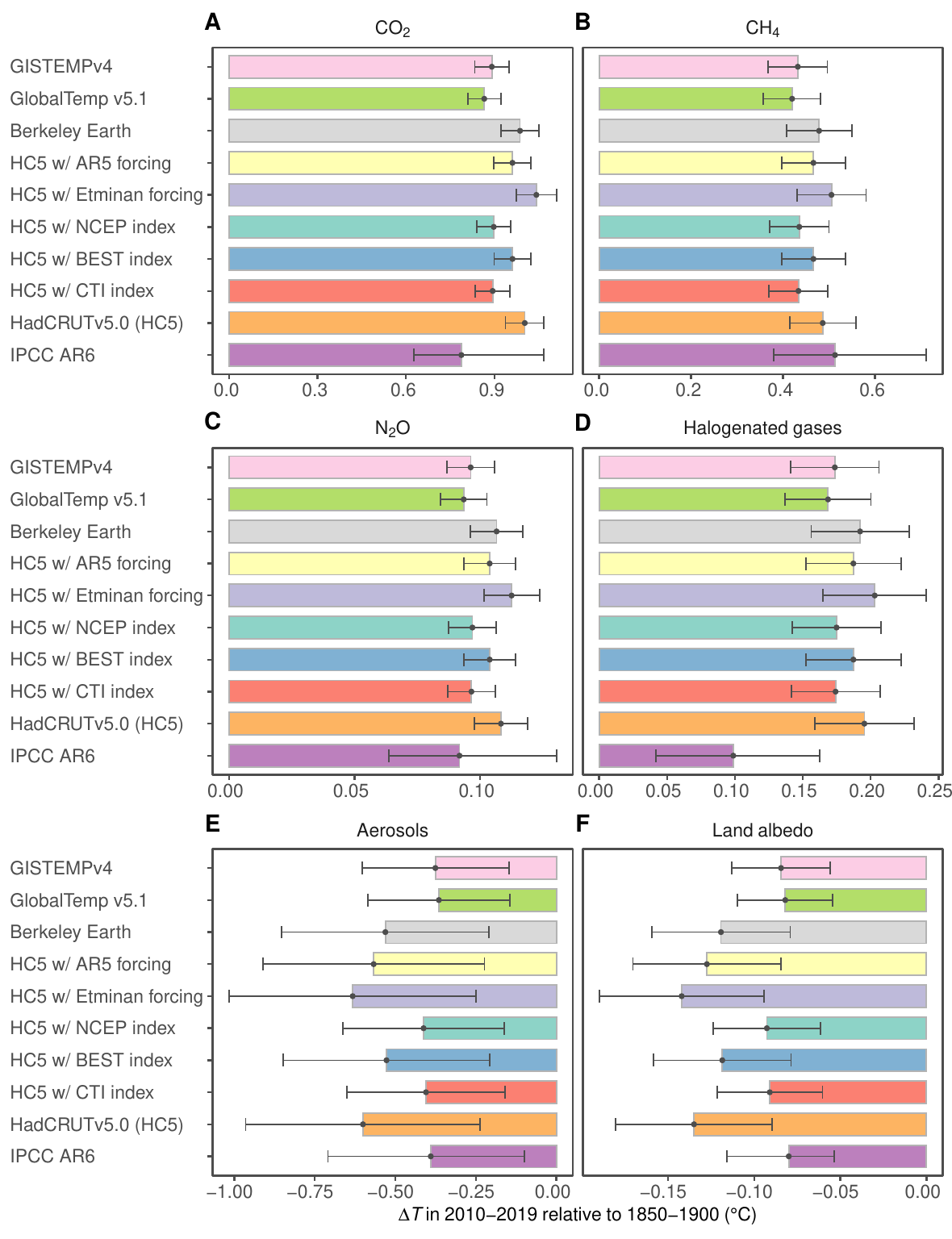}
    \caption{\textbf{Attributions of \boldmath{$\Delta T$} in this study vs. AR6.} The AR6 attribution was extracted from Figure SPM.2. Error bars in both this study and AR6 are one-sigma uncertainties. Note that Figure SPM.2 shows very likely ranges (5\textsuperscript{th}-95\textsuperscript{th} percentiles). We estimated the one-sigma uncertainty based on the AR6 calculation (\protect\url{https://github.com/sarambl/AR6_CH6_RCMIPFIGS}) (see materials and methods).}
    \label{fig:compare2}
\end{figure}

To estimate the uncertainty of aerosols presented in fig. \ref{fig:compare2}, we summed the aerosols' \textsl{$\Delta T$} from AR6, namely, NO\textsubscript{x}, VOC, SO\textsubscript{2}, OC, BC and NH\textsubscript{3}. Their uncertainties were propagated by assuming that the individual \textsl{$\Delta T$} values are independent variables that are normally distributed. Thus, their sum was also normally distributed. For example, from $X\sim N(\mu_{X},\sigma_{X}^{2})$ and $Y\sim N(\mu_{Y},\sigma_{Y}^{2})$, we obtained the uncertainty of $Z=X+Y$: $Z\sim N(\mu _{X}+\mu _{Y},\sigma _{X}^{2}+\sigma _{Y}^{2})$. A similar estimation was used in the AR6 calculation (\url{https://github.com/sarambl/AR6_CH6_RCMIPFIGS}).

\clearpage
\pagebreak

\section*{Supplementary Figures}

\clearpage
\pagebreak

\section*{Supplementary Tables}

\begin{table}[ht]
    \centering
    \caption{Test of statistical models using the Bayesian Information Criterion (BIC)}
    \label{tab:bic}
    \begin{tabular}{p{6mm}p{40mm}p{28mm}p{50mm}}
    \hline
    Index & Regression among variables                                                         & BIC & Remark                                                                                \\
    \hline
    1     & GHGs, Non-GHGs, natural, MEI                                                       & -1216.766                           & Used in this study                                                                    \\
    2     & Total sum, MEI                                                                     & -1180.292                           &                                                                                       \\
    3     & Anthropogenic, natural, MEI                                                        & -1205.087                           &                                                                                       \\
    4     & GHGs, aerosols, land albedo, natural, MEI                                          & -1211.329                           & Negative coefficient occurs in land albedo, invalid                                 \\
    5     & GHGs, Non-GHGs, volcanic, solar, MEI                                               & -1216.033                           &                                                                                       \\
    6     & CO\textsubscript{2}, CH\textsubscript{4}, N\textsubscript{2}O, ODS, other fluorinated, aerosols, land albedo, natural, MEI        & -1206.71                            & Negative coefficients occur in CH\textsubscript{4}, N\textsubscript{2}O, other fluorinated and land albedo, invalid \\
    7     & CO\textsubscript{2}, CH\textsubscript{4}, N\textsubscript{2}O, ODS, other fluorinated, aerosols, land albedo, volcanic, sola, MEI & -1225.454                           & Negative coefficients occur in CH\textsubscript{4}, N\textsubscript{2}O and land albedo, invalid \\
    \hline                  
    \end{tabular}
\end{table}

\begin{table}[ht]
    \scriptsize
    \centering
    \caption{List of halogenated compounds used in this study and observational records}
    \label{tab:comp}
    \begin{tabular}{p{20mm}p{20mm}p{20mm}p{20mm}p{20mm}p{20mm}}
    \hline
    Compounds     & This study & CMIP6 & NOAA & AGAGE\_md (*) & AGAGE\_ms (\textdagger) \\
    \hline
    \multicolumn{6}{l}{Ozone-Depleting Substances (ODS)}\\
    \hline
    CFC-11       & $\surd$          & $\surd$     & $\surd$    & $\surd$         &           \\
    CFC-12       & $\surd$          & $\surd$     & $\surd$    & $\surd$         &           \\
    CFC-13       &                      &               &               &          & $\surd$          \\
    CFC-113      & $\surd$          & $\surd$     & $\surd$    & $\surd$         & $\surd$         \\
    CFC-114      & $\surd$          & $\surd$     &      &           & $\surd$         \\
    CFC-115      & $\surd$          & $\surd$     &      &           & $\surd$         \\
    HCFC-22      & $\surd$          & $\surd$     & $\surd$    &           & $\surd$         \\
    HCFC-141b    & $\surd$          & $\surd$     & $\surd$    &           & $\surd$         \\
    HCFC-142b    & $\surd$          & $\surd$     & $\surd$    &           & $\surd$         \\
    CH\textsubscript{3}CCl\textsubscript{3}      & $\surd$          & $\surd$     & $\surd$    & $\surd$         & $\surd$         \\
    CCl\textsubscript{4}         & $\surd$          & $\surd$     & $\surd$    & $\surd$         &           \\
    CH\textsubscript{3}Cl        & $\surd$          & $\surd$     &      &           & $\surd$         \\
    CH\textsubscript{2}Cl\textsubscript{2}       &            & $\surd$     &      &           & $\surd$         \\
    CHCl\textsubscript{3}        &            & $\surd$     &      & $\surd$         & $\surd$         \\
    CH\textsubscript{3}Br        & $\surd$          & $\surd$     & $\surd$    &           & $\surd$         \\
    Halon-1211   & $\surd$          & $\surd$     & $\surd$    &           & $\surd$         \\
    Halon-1202   & $\surd$          &       &      &           &           \\
    Halon-1301   & $\surd$          & $\surd$     & $\surd$    &           & $\surd$         \\
    Halon-2402   & $\surd$          & $\surd$     & $\surd$    &           & $\surd$         \\
    \hline
    \multicolumn{6}{l}{Other fluorinated gases}\\
    \hline
    HFC-134      & $\surd$          &       &      &           &           \\
    HFC-134a     & $\surd$          & $\surd$     & $\surd$    &           & $\surd$         \\
    HFC-23       & $\surd$          & $\surd$     &      &           & $\surd$         \\
    HFC-32       & $\surd$          & $\surd$     & $\surd$    &           & $\surd$         \\
    HFC-125      & $\surd$          & $\surd$     & $\surd$    &           & $\surd$         \\
    HFC-143      & $\surd$          &       &      &           &           \\
    HFC-143a     & $\surd$          & $\surd$     & $\surd$    &           & $\surd$         \\
    HFC-152a     & $\surd$          & $\surd$     & $\surd$    &           & $\surd$         \\
    HFC-227ea    & $\surd$          & $\surd$     & $\surd$    &           & $\surd$         \\
    HFC-236fa    & $\surd$          & $\surd$     &      &           & $\surd$         \\
    HFC-245fa    & $\surd$          & $\surd$     &      &           & $\surd$         \\
    HFC-365mfc   & $\surd$          & $\surd$     & $\surd$    &           & $\surd$         \\
    HFC-41       & $\surd$          &       &      &           &           \\
    HFC-43-10mee & $\surd$          & $\surd$     &      &           & $\surd$         \\
    NF\textsubscript{3}          & $\surd$          & $\surd$     &      &           & $\surd$         \\
    SF\textsubscript{6}          & $\surd$          & $\surd$     &      &           & $\surd$         \\
    SO\textsubscript{2}F\textsubscript{2}        &            & $\surd$     &      &           & $\surd$         \\
    CF\textsubscript{4}          & $\surd$          & $\surd$     &      &           & $\surd$         \\
    C\textsubscript{2}F\textsubscript{6}         & $\surd$          & $\surd$     &      &           & $\surd$         \\
    C\textsubscript{3}F\textsubscript{8}         & $\surd$          & $\surd$     &      &           & $\surd$         \\
    C\textsubscript{4}F\textsubscript{10}        & $\surd$          & $\surd$     &      &           &           \\
    C\textsubscript{5}F\textsubscript{12}        & $\surd$          & $\surd$     &      &           &           \\
    C\textsubscript{6}F\textsubscript{14}        & $\surd$          & $\surd$     &      &           &           \\
    C\textsubscript{7}F\textsubscript{16}        &            & $\surd$     &      &           &           \\
    C\textsubscript{8}F\textsubscript{18}        &            & $\surd$     &      &           &           \\
    c-C\textsubscript{4}F\textsubscript{8}       & $\surd$          & $\surd$     &      &           &           \\
    C\textsubscript{2}Cl\textsubscript{4}        &                      &               &               &          & $\surd$          \\
    \hline  
    \multicolumn{6}{l}{\textsuperscript{*} AGAGE\_md: AGAGE GC-MD data}\\
    \multicolumn{6}{l}{ (\protect\url{https://agage2.eas.gatech.edu/data_archive/global_mean/global_mean_md.txt})}\\
    \multicolumn{6}{l}{\textsuperscript{\textdagger} AGAGE\_ms: AGAGE GCMS-Medusa data}\\
    \multicolumn{6}{l}{ (\protect\url{https://agage2.eas.gatech.edu/data_archive/global_mean/global_mean_ms.txt})}
    \end{tabular}
\end{table}

\begin{table}[ht]
    \centering
    \caption{The investigated CMIP6 ESMs and ensemble sizes}
    \label{tab:cmip6}
    \begin{tabular}{p{1.5cm}p{4cm}p{2cm}p{2cm}p{2cm}}
    \hline
    Index & Model           & hist-GHG & hist-aer & hist-nat \\
    \hline
    1     & ACCESS-CM2      & 3        & 3        & 3        \\
    2     & ACCESS-ESM1-5   & 3        & 3        & 3        \\
    3     & BCC-CSM2-MR     & 3        & 3        & 3        \\
    4     & CanESM5         & 50       & 30       & 50       \\
    5     & CESM2           & 3        & 2        & 3        \\
    6     & CNRM-CM6-1      & 10       & 10       & 10       \\
    7     & FGOALS-g3       & 3        & 3        & 3        \\
    8     & GFDL-CM4        & 0        & 0        & 3        \\
    9     & GFDL-ESM4       & 1        & 1        & 3        \\
    10    & GISS-E2-1-G     & 10       & 15       & 20       \\
    11    & HadGEM3-GC31-LL & 5        & 5        & 10       \\
    12    & IPSL-CM6A-LR    & 10       & 10       & 10       \\
    13    & MIROC6          & 3        & 10       & 50       \\
    14    & MRI-ESM2-0      & 5        & 5        & 5        \\
    15    & NorESM2-LM      & 3        & 3        & 3        \\
    \hline
    \end{tabular}
    \end{table}

    \begin{table}[ht]
        \small
        \caption{Fitted coefficients for the statistical model}
        \label{tab:coef}
        \begin{tabular}{p{23mm}p{20mm}p{20mm}p{23mm}p{23mm}p{16mm}}
        \hline
        Corresponding figures and estimates & Berkeley Earth   & GISTEMPv4         & GlobalTemp v5.1  & HadCRUT5 (mainly used) &  HadCRUT4.6 \\
        \hline 
        In Figs.            & figs. \ref{fig:compare2}, \ref{fig:attrib_berk}                & figs. \ref{fig:compare2}, \ref{fig:attrib_gis}                & figs. \ref{fig:compare2}, \ref{fig:attrib_glb}                & Figs. \ref{fig:demo}, \ref{fig:attrib}, \ref{fig:cause}, \ref{fig:nat} \newline figs. \ref{fig:compare2}, \ref{fig:hist_hd50}, \ref{fig:trd_ghg}, \ref{fig:dif_regre_hd50}, \ref{fig:conc_trd}, \ref{fig:trd_aer}, \ref{fig:vari}  &  fig. \ref{fig:attrib_hd46}                              \\
        $\beta_{GHG}$       & 1.28$\pm$0.20   & 1.16$\pm$0.15   & 1.13$\pm$0.14   & 1.31$\pm$0.22     &   1.08$\pm$0.22 \\
        $\beta_{non-GHG}$   & 1.83$\pm$0.70   & 1.30$\pm$0.48   & 1.26$\pm$ 0.48  & 2.07$\pm$0.81     &   1.57$\pm$0.84 \\
        $\beta_{nat}$       & 1.51$\pm$0.19   & 1.52$\pm$0.19   & 1.51$\pm$0.19   & 1.56$\pm$0.21     &   1.32$\pm$0.32 \\
        $\beta^{1}_{MEI}$   & 0.048$\pm$0.001 & 0.017$\pm$0.001 & 0.039$\pm$0.001 & 0.057$\pm$0.002   &    0.062$\pm$0.001\\
        $\beta^{2}_{MEI}$   & 0.016$\pm$0.000 & 0.017$\pm$0.001 & 0.000$\pm$0.000 & 0.006$\pm$0.002   &    0.028$\pm$0.002\\
        $\beta^{3}_{MEI}$   & 0.043$\pm$0.001 & 0.038$\pm$0.000 & 0.039$\pm$0.000 & 0.041$\pm$0.002   &    0.025$\pm$0.002\\
        $\tau_{1}$   & 0               & 2               & 2               & 0                        &     0\\
        $\tau_{2}$   & 1               & 2               & 3               & 6                        &     7\\
        $\tau_{3}$   & 9               & 3               & 4               & 8                        &     10\\
        Mean $R^{2}$        & 0.87            & 0.87            & 0.88            & 0.88              &     0.86\\
        Mean correl. between ENSO \& residuals       & 2.8E-17           & 6.8E-18           & 8.7E-17           & 3.2E-17   &  1.5E-17      \\
        Mean correl. between estimate \& observation       & 0.93           & 0.93          & 0.94          & 0.94          &   0.93 \\
        \hline   
        \end{tabular}
    \end{table}
    
    \begin{table}[ht]
        \small
        \centering
        \renewcommand\thetable{}
        \caption{Fitted coefficients for the statistical model (continued)}
        \begin{tabular}{p{23mm}p{20mm}p{20mm}p{20mm}p{21mm}p{20mm}}
        \hline
        Corresponding figures and estimates       & HadCRUT5 w/ NCEP index & HadCRUT5 w/ CTI  & HadCRUT5 w/ BEST index & HadCRUT5 w/ Etminan forcing   & HadCRUT5 w/ AR5 forcing \\
        \hline 
        In Figs.             & figs. \ref{fig:compare2}, \ref{fig:vari_ncep}, \ref{fig:attrib_hd50_ncep}                & figs. \ref{fig:compare2}, \ref{fig:vari_cti}, \ref{fig:attrib_hd50_cti}                & figs. \ref{fig:compare2}, \ref{fig:vari_best}, \ref{fig:attrib_hd50_best}                & figs. \ref{fig:compare2}, \ref{fig:attrib_et_hd50}                &  figs. \ref{fig:compare2}, \ref{fig:attrib_my_hd50}               \\
        $\beta_{GHG}$        & 1.17$\pm$0.17    & 1.16$\pm$0.19   & 1.25$\pm$0.19   & 1.36$\pm$0.23   & 1.25$\pm$ 0.20         \\
        $\beta_{non-GHG}$    & 1.42$\pm$0.69   & 1.40$\pm$0.70   & 1.82$\pm$0.73   & 2.18$\pm$0.82   & 1.96$\pm$ 0.80          \\
        $\beta_{nat}$        & 0.84$\pm$0.30   & 0.91$\pm$0.34   & 1.66$\pm$0.22   & 1.64$\pm$0.21   & 1.49$\pm$0.18          \\
        $\beta^{1}_{MEI}$    & 0.035$\pm$0.002 & 0.068$\pm$0.003 & 0.054$\pm$0.002 & 0.057$\pm$0.002 & 0.057$\pm$0.002         \\
        $\beta^{2}_{MEI}$    & 0.016$\pm$0.002 & 0.032$\pm$0.003 & 0.029$\pm$0.002 & 0.006$\pm$0.002 & 0.006$\pm$0.002         \\
        $\beta^{3}_{MEI}$    & 0.021$\pm$0.002 & 0.035$\pm$0.003 & 0.018$\pm$0.002 & 0.040$\pm$0.002 & 0.041$\pm$ 0.003        \\
        $\tau_{1}$    & 6               & 3               & 0               & 0               & 0                         \\
        $\tau_{2}$    & 7               & 8               & 8               & 6               & 6                         \\
        $\tau_{3}$    & 8               & 11              & 10              & 8               & 8                         \\
        Mean $R^{2}$         & 0.87            & 0.84            & 0.86            & 0.88            & 0.88                      \\
        Mean correl. between ENSO \& residuals       & 1.3E-17           & -8.7E-18          & -6.2E-18          & 6.7E-18           & 8.5E-18     \\
        Mean correl. between estimate \& observation       & 0.93           & 0.92          & 0.92          & 0.94           & 0.94     \\
        \hline   
        \end{tabular}
    \end{table}

\pagebreak
\clearpage



\end{document}